\documentclass[12pt]{article}
\usepackage{amssymb,amsmath,amscd,epsfig}
\usepackage{psfrag}
\makeatletter

\@addtoreset{equation}{section}
\def\section{\@startsection {section}{1}{\z@}{-2.25ex plus -1ex minus
 -.2ex}{1.0ex plus .2ex}{\large\bf}}
\def\subsection{\@startsection{subsection}{2}{\z@}{-2.0ex plus%
 -1ex minus -.2ex}{0.5ex plus .2ex}{\bf}}

\def\Ad{\mathrm{Ad}}
\def\ad{\mathrm{ad}}

\newcommand{\inv}[0]{{-1}}

\newcommand{\cif}[0]{\mathcal{C}^\infty}

\def\bel{{\mbox{\boldmath $\ell$}}}
\def\bv{{\mbox{\boldmath $v$}}}

\def\baa{{\mbox{\boldmath $a$}}}
\def\bx{{\mbox{\boldmath $x$}}}
\def\bX{{\mbox{\boldmath $X$}}}
\def\bs{{\mbox{\boldmath $s$}}}

\def\bj{{\mbox{\boldmath $j$}}}

\def\bp{{\mbox{\boldmath $p$}}}

\def\bq{{\mbox{\boldmath $q$}}}

\def\bJ{{\mbox{\boldmath $J$}}}

\newcommand{\gothg}{\mathfrak g }

\newcommand{\gothh}{\mathfrak h }

\newcommand{\ZZ}{\mathbb{Z}}

\newcommand{\RR}{\mathbb{R}}
\newcommand{\CC}{\mathbb{C}}

\newtheorem{theorem}{Theorem}[section]

\newtheorem{example}[theorem]{Example}

\def\bmz{\left(\begin{array}{2,2}}
\def\emz{\end{array}\right)}
\def\bmd{\left(\begin{array}{3,3}}
\def\emd{\end{array}\right)}
\newcommand{\be}{\begin{equation}}
\newcommand{\ee}{\end{equation}}
\newcommand{\ba}{\begin{eqnarray}}
\newcommand{\ea}{\end{eqnarray}}

\def\bpm{\begin{pmatrix}}
\def\epm{\end{pmatrix}}

\begin{document}
\parskip 6pt
\parindent 0pt
\begin{flushright}
pi-qg-93\\
\end{flushright}

\begin{center}
\baselineskip 24 pt {\large \bf   The Hilbert space of 3d gravity:
quantum group symmetries and
observables 
}

\baselineskip 16 pt

\vspace{.7cm} { C.~Meusburger}\footnote{\tt  catherine.meusburger@math.uni-erlangen.de }\\
Department Mathematik, Friedrich-Alexander-Universit\"at Erlangen-N\"urnberg,\\
Cauerstra\ss e 11,  91058 Erlangen, \\
Germany\\

\vspace{.5cm}
{ K.~Noui}\footnote{\tt karim.noui@lmpt.univ-tours.fr} \\
Laboratoire de Math\'ematiques et de Physique Th\'eorique \\
F\'ed\'eration Denis Poisson Orl\'eans-Tours, CNRS/UMR 6083 \\
Facult\'e des Sciences, Parc de Grammont, 37200 Tours, France \\

\vspace{0.5cm}

{ February 5 2012}

\end{center}

\begin{abstract}
\noindent  
We relate three-dimensional loop quantum gravity to the combinatorial quantisation formalism 
based on the Chern-Simons formulation for three-dimensional Lorentzian and Euclidean gravity with vanishing cosmological constant. We compare  the construction of the kinematical Hilbert space and the implementation of the constraints. This leads to an explicit and very interesting
relation between the associated operators in the two approaches and sheds light on their physical interpretation.  We demonstrate that the quantum group symmetries arising in the combinatorial formalism, the quantum double of the three-dimensional Lorentz and rotation group, are also present in the loop formalism. We derive
 explicit expressions for the action of these quantum groups  on the space of cylindrical functions associated with graphs. This establishes a direct link between the two quantisation approaches and clarifies the role of quantum group symmetries in three-dimensional gravity.

 \end{abstract}


\section{Introduction}

\subsection{Motivation}

One of the main motivations for the study of three-dimensional gravity is its role as a toy model for quantum gravity. It allows one to investigate conceptual questions of quantum gravity, serves as 
a testing ground for quantisation formalisms and has inspired approaches for the four-dimensional case.
This is due to the fact that Einstein's theory of gravity simplifies significantly in three dimensions: It has no local gravitational degrees of freedom, but a finite number of global degrees of freedom arising for spacetimes with non-trivial topology or with point particles. As the phase space of the theory is finite dimensional, its quantisation simplifies considerably compared to the four-dimensional case. Important progress towards quantisation has been achieved within
many approaches, for an overview see \cite{Carlipbook}. As in higher dimensions, two of the most prominent ones are loop quantum gravity and spin-foam models. 
Further progress followed the discovery that three-dimensional gravity can be formulated as a Chern-Simons 
gauge theory \cite{AT, Witten1}. 

The Chern-Simons formulation of the theory gave rise to important advances on the conceptual level as well as an improved understanding of the mathematical structure of the theory. In particular, it relates the phase space of the theory to moduli spaces of flat connections on two-dimensional surfaces and establishes a relation with the theory of knot invariants \cite{Witten2} and manifold invariants \cite{TuVir}. It also lead to the development of new and powerful quantisation approaches.

\subsubsection{Combinatorial Quantisation and the loop formalism}

One of these approaches  which will play a central role in this paper is the combinatorial quantisation formalism for Chern-Simons gauge theory. This formalism, first established in \cite{AGSI,AGSII,AS,BR} for for Chern-Simons theories with compact, semisimple gauge groups, has been generalised to the gauge groups arising in three-dimensional gravity in \cite{BNR, we2}. It lead to important advances in the quantisation of the theory, specifically in the construction of the physical Hilbert space.  Moreover, it 
provides powerful mathematical tools, namely the theory of Hopf algebras and quantum groups,  which arise naturally in this formalism.

Despite these advances, 
many important issues related to the quantisation of three-dimensional gravity  remain to be resolved: It is currently not clear how different quantisation formalisms for the theory are related and if they lead to equivalent quantum theories. This question is especially relevant for the relation between three-dimensional loop quantum gravity and the combinatorial quantisation formalism, as these approaches follow a very similar quantisation philosophy. Both pursue a Hamiltonian quantisation approach, they are
 based on a (2+1)-decomposition of the underlying manifold, and their fundamental variables are holonomies associated to graphs on the two-dimensional spatial surface.  
 
 This suggests that the link between three-dimensional loop quantum gravity and the combinatorial quantisation formalism should be direct,  and that it should be possible to explicitly relate the resulting quantum theories. Moreover,  the main conceptual difference between these approaches is that they are based, respectively, on the BF and the Chern-Simons formulation of the theory. Understanding the relation between these approaches would therefore not only contribute to the understanding of three-dimensional quantum gravity itself but also shed light on issues surrounding the relation between three-dimensional gravity and Chern-Simons theory. 
 
However, despite its relevance and its conceptual importance,  the relation between these two quantisation approaches is currently not well-understood.  Its clarification is one of the core results of this paper.  In the following, we explicitly relate the construction of their kinematical and physical Hilbert spaces. Moreover, we demonstrate how the associated quantum operators in the combinatorial formalism can be expressed in terms of the operators in loop quantum gravity and that the link between these variables has a clear physical interpretation.

\subsubsection{Quantum group symmetries}

The other central result of our paper addresses the role of quantum group symmetries in the two approaches. 
As powerful mathematical tools, they are of practical relevance for the quantisation of the theory.
However,  quantum groups and, more generally, Hopf algebras are also discussed as generic symmetries of quantum gravity and believed to reflect fundamental properties of quantum  spacetimes.  The idea is that  spacetimes  loose their smoothness near the Planck scale and instead acquire a fuzzy, discrete or non-commutative structure. 
It has been argued that this corresponds to a deformation of their
local symmetry groups into a Hopf algebra symmetries. 
Although such deformations via Hopf algebras have been investigated extensively \cite{Mad,DoNe,Chai,Jihad,JMN}, 
their status in four dimensions remains largely heuristic due to the difficulties in the quantisation of the theory.

In three-dimensional gravity, the situation is less involved and can be investigated with more rigour. Quantum groups arise naturally  in the combinatorial quantisation  formalisms \cite{BNR,we2} but also in other approaches \cite{TuVir}. For three-dimensional gravity with vanishing cosmological constant, the relevant quantum groups are the quantum (or Drinfeld) doubles $D(G)$, where, depending on the signature, $G$ is the three-dimensional rotation group $SU(2)$ or the three-dimensional Lorentz group $SU(1,1)\cong SL(2,\RR)$. They are deformations of the local isometry groups of the classical spacetimes, respectively, the three-dimensional Euclidean and Poincar\'e group. The deformation parameter is the Planck length
$\ell_P=\hbar G_N$, where  $G_N$ is the Newton constant in three dimensions.
Classical observables, which are (by definition) invariant under  these classical symmetry groups become quantum observables which form an algebra and  are invariant under the action of the quantum double
$D(G)$. 

Although quantum groups arise in the combinatorial quantisation of Euclidean and Lorentzian three-dimensional gravity with vanishing cosmological constant \cite{we2}, they are not readily apparent in
 three-dimensional loop quantum gravity and  in the Ponzano-Regge model \cite{prmodel}. The relation between the Ponzano-Regge model and the evaluation of link invariants for the quantum double $D(SU(2))$ has been investigated in \cite{LPR}, but only specific representations of $D(SU(2))$ are considered and the role of quantum group symmetries remains implicit. For a more recent result concerning the mathematical structure and the role of  link invariants in the Ponzano-Regge model  see \cite{BPR}.
This absence of quantum group symmetries in the loop and spin foam formalisms raised the question if they are a generic feature of three-dimensional quantum gravity or merely a tool limited to the combinatorial quantisation formalism.

In this paper we show that quantum group symmetries are a generic feature of three-dimensional gravity with vanishing cosmological constant and that they are also present in three-dimensional loop quantum gravity. We demonstrate that the quantum doubles $D(SU(2))$ and $D(SU(1,1))$ act naturally on the Hilbert spaces of the theory, i.~e.~the space of cylindrical functions associated with graphs. As the cylindrical functions are closely related to the spin network functions which are the fundamental building blocks of the quantum theory in loop quantum gravity and the spinfoam approach, this establishes the presence of quantum group symmetries in these formalisms. 
We show that each closed, non-selfintersecting loop in the graph gives rise to a representation of the quantum double on the space of cylindrical functions and derive explicit expressions for these representations. Moreover, we demonstrate that these 
representations are intimately related to the implementation of the constraints in the quantum theory.

\subsection{Outline of the paper}

Our paper is structured as follows:  In Sect.~\ref{classect}  we summarise and contrast  the classical formulations of the theory 
underlying 3d loop quantum gravity and the combinatorial quantisation formalism. These are, respectively,  the BF formulation and the Chern-Simons formulation of three-dimensional gravity with vanishing cosmological constant. We review the canonical analysis in the two formulations and discuss their gauge and physical symmetries. 

In Sect.~\ref{discretesect}, we give a detailed discussion of the discretisation of the phase space which serves as the starting point for the two quantisation approaches.  In both approaches, this discretisation is based on a graph embedded in the spatial surface, in case of the combinatorial formalism, equipped with additional structure \cite{FR}. We summarise the construction of the discrete phase space variables and their Poisson structure as well as implementation of the constraints and the description of the physical phase space. This discussion motivates the different quantisation approaches and lays the foundation for the following sections in which we relate the associated quantum theories.

In Sect.~\ref{hilbsect} we relate the associated quantum theories. In both formalisms the quantum  states are cylindrical functions based on a graph. However, the operators which act on these spaces differ, and  there is a priori no direct link between the fundamental variables in the two approaches.  The core result of this section is an explicit formula relating the quantum operators in the loop and the combinatorial formalism. Moreover, we show that this relation has a clear physical interpretation and that it sheds light on the role of the additional structures present in the combinatorial quantisation formalism.

Sect.~\ref{qudoublesymms} is concerned with the other central aspect of our paper,  the role of quantum group symmetries. We show that the quantum doubles of the three-dimensional rotation and Lorentz group arise naturally not only in the combinatorial formalism but also in three-dimensional loop quantum gravity. More specifically, we demonstrate that each non-selfintersecting loop in the underlying graph gives rise to a representation of the quantum double on the associated space of cylindrical functions. 
This establishes and clarifies the role of quantum groups in three-dimensional quantum gravity.
Moreover, we find that these quantum group symmetries have a natural interpretation and play an important role in the construction of the kinematical and physical Hilbert space.

The construction of the physical Hilbert space and the implementation of the constraints in the two quantisation formalism are the subject of 
Sect.~\ref{physhilb}. We  show that the standard gauge fixing procedure via 
contractions of maximal trees has a natural interpretation in the combinatorial formalism which arises from the classical graph operations defined by Fock and Rosly \cite{FR}. Moreover, we demonstrate that the implementation of the constraints is closely related to the representations of the quantum double in Sect.~\ref{qudoublesymms}, which unify the requirements of graph gauge invariance and the projector on the physical Hilbert space.   

Sect.~\ref{concsect} contains our conclusions and outlook. Appendix \ref{frphsp} summarises the formalism of Fock and Rosly  \cite{FR} and its application to the phase space of three-dimensional gravity. Appendix \ref{qurep}  presents some relevant facts from the representation theory of the quantum doubles $D(SU(2))$, $D(SU(1,1))$.

\section{Classical 3d gravity in the BF formulation and the Chern-Simons formulation}
\label{classect}

\subsection{Definitions and notation}
In this paper, we consider three-dimensional gravity of 
Euclidean and Lorentzian signature and with vanishing cosmological constant. 
We introduce a ``space-time" manifold  ${\cal M}$.  Through most of the paper we assume it to be of topology $M\approx S\times \mathbb I$ where the spatial surface $S$ is an orientable two-surface of general genus and, possibly, with punctures representing massive point particles. The interval $\mathbb I\subset \RR$ characterises the ``time" direction.

We choose a local coordinate system $(x_\mu)_{\mu=0,1,2}$ of $\cal M$. In the following, Greek letters $\mu,\nu,\cdots$ refer to 
space-time indices, Latin letters $i,j,\cdots$ to space indices, and $t$ is the time index. 
 Latin letters  $a,b,\cdots$ from the beginning of the alphabet stand for indices associated with  Lie groups and Lie algebras. Throughout the paper we use Einstein's summation convention. 
Indices are raised and lowered with either the three-dimensional Minkowski metric $\text{diag}(1,-1,-1)$ or the three-dimensional Euclidean metric $\text{diag}(1,1,1)$, both of which are denoted by $\eta$.  With that convention, all formulas refer to both Lorentzian and Euclidean signature unless specified otherwise.

Throughout the paper, we write $G$ for both  the three-dimensional rotation group  $G=SU(2)$ and 
the three-dimensional Lorentz group
 $G=SU(1,1)$.  We fix a set  of generators  $J_a$, $a=0,1,2$,  of their Lie algebras $\gothg=\text{Lie}\,G$ in terms of  which the Lie bracket takes the form
  \begin{align}
  \label{gbrack}
[J_a,J_b]=\epsilon_{abc}J^c.
\end{align}
Here,  $\epsilon$ is the totally anti-symmetric tensor in three dimensions with the convention $\epsilon_{012}=1$ and indices are raised and lowered with the three-dimensional Minkowski and Euclidean metric. We denote by $\Ad$ the adjoint action of $G$ on its Lie algebra $\gothg\cong\RR^3$
\begin{align}
u\cdot (v^aJ_a)\cdot u^\inv=\Ad(u)^b_{\;\;a}v^a J_b\qquad \forall u\in G,\bv\in\RR^3.
\end{align}
We also introduce the left-  and right  invariant vector fields  $L^a$ and $R^a$
on $G$,
\begin{eqnarray}
\label{vecfieldact}
R^af(g) & = & df(R^a)=\frac{d}{dt}|_{t=0} f(g\cdot e^{tJ_a})\\ 
L^af(g) & = & df(L^a)=\frac{d}{dt}|_{t=0} f(e^{-tJ_a}\cdot g) \qquad \forall g\in G, f\in\cif(G)\;.\nonumber
\end{eqnarray}

The local symmetry groups of  Euclidean and Lorentzian  (2+1)-gravity with vanishing cosmological constant are, respectively,  the three-dimensional Euclidean group  and the three-dimensional Poincar\'e group. They have the structure of a  semidirect product  $G\ltimes\RR^3$ and will be denoted by $IG$ in the following.
With the parametrisation
\begin{align}
\label{gparam}
(u,\baa)=(u,-\Ad(u)\bj)\qquad u\in G,\; \bj,\baa \in \RR^3
\end{align}
their group multiplication law reads 
\begin{align}
\label{gmult}
(u_1,\baa_1)\cdot(u_2,\baa_2)=(u_1u_2,\baa_1+\Ad(u_1)\baa_2)
.
\end{align}
The associated Lie algebras $\mathfrak g \ltimes \RR^3$ are parametrised by the generators $J_a$, $a=0,1,2$, and an additional set of generators  $P_a$, $a=0,1,2$, which correspond to the infinitesimal translations. In terms of these generators, the Lie bracket takes the form
\begin{align}
\label{poincbrack}
[J_a,J_b]=\epsilon_{abc} J^c\qquad
[J_a,P_b]=\epsilon_{abc} P^c\qquad 
[P_a,P_b]=0,
\end{align}
and an $\Ad$-invariant, non-degenerate symmetric bilinear 
form on $\mathfrak g \ltimes \RR^3$ is given by 
\begin{align}
\label{sblf}
\langle J_a,J_b\rangle=0\qquad
\langle J_a,P_b\rangle=\eta_{ab}\qquad 
\langle P_a,P_b\rangle=0.
\end{align}

\subsection{Classical gravity in three dimensions}

\subsubsection{First order gravity:  the BF formulation and the Chern-Simons formulation}

It is well-known that solutions of pure general relativity in three dimensions are locally trivial. 
This particularity is manifest when one writes the pure gravity action in the first order formalism,
where the dynamical variables are $\mathfrak g$-valued one-forms: the triad $e=e_\mu^aJ_adx^\mu$ 
which defines the metric via
\begin{equation}
\label{metricdef}
g_{\mu\nu}=e_\mu^ae_\nu^b\eta_{ab}
\end{equation}
and the spin-connection $\omega=\omega_\mu^aJ_adx^\mu$, which is closely related to the Levi-Civita connection. 
When expressed in terms of these variables, the Einstein-Hilbert action reduces to a topological $BF$ type action
\begin{eqnarray}\label{first order action}
S_{BF}[e,\omega] \; = \; \alpha  \, \int_{\cal M} d^3x \, \epsilon^{\mu\nu\rho}\eta_{ab} \,e_\mu^a \, F_{\nu\rho}^b[\omega],
\end{eqnarray}
where  $\alpha=(4\pi  G_N)^{-1}$ is related to the three dimensional Newton constant $G_N$ and 
will be set to one in the following.   $F_{\mu \nu}[\omega]$ is the curvature of the G-connection $\omega$
\begin{align}
\label{curvature}
F_{\mu \nu}[\omega]=\partial_\mu \omega_\nu -  \partial_\nu \omega_\mu +\tfrac 1 2  [\omega_\mu, \omega_\nu].
\end{align}
In fact, the first order formulation of (2+1)-gravity gives rise to two equivalent formulations of the classical theory, the BF formulation above which underlies three-dimensional loop quantum gravity and the formulation as a  Chern-Simons gauge theory which is the starting point for the combinatorial quantisation formalism.
To obtain the Chern-Simons formulation of the theory, one combines triad and spin connection into a Chern-Simons gauge  field
\begin{align} 
A=e^aP_a + \omega^aJ_a,
\end{align} 
which is a one-form with values in the three-dimensional Poincar\'e or Euclidean algebra ${\mathfrak g}\ltimes \mathbb R^3$.
It is shown in \cite{AT, Witten1} that the first order action for three-dimensional gravity  can then be rewritten as a Chern-Simons 
action 
\begin{eqnarray}\label{CS action}
S_{CS}[A(e,\omega)] & = & \int_{\cal M} d^3x \, \epsilon^{\mu \nu \rho}
(\langle A_\mu, \partial_\nu A_\rho \rangle + \frac{1}{3} \langle A_\mu, [A_\nu, A_\rho] \rangle) 
\end{eqnarray}
where $\langle,\rangle$ is the  bilinear form  (\ref{sblf}).  
Using the formula for the Lie bracket \eqref{poincbrack}, it is easy to check that this action is equivalent to \eqref{first order action}  up to a boundary term for $\partial M\neq \emptyset$, which does not modify the equations of motion.  

Varying  the actions \eqref{first order action}, \eqref{CS action} with respect to the triad and spin connection results in a flatness condition on the $IG$-valued Chern-Simons connection $A$. This flatness condition combines  the requirements of flatness for the spin connection $\omega$ and of vanishing torsion (i.~e.~the requirement that the triad $e$ is covariantly constant with respect to $\omega$)
\begin{eqnarray}
\label{flatcond}
F_{\mu\nu}[A] \; = \; 0 \; \Longleftrightarrow \; \begin{cases} 
 &F_{\mu \nu}[\omega] \equiv \partial_\mu\omega_\nu - \partial_\nu \omega_\mu  + \tfrac 1 2 [\omega_\mu, \omega_\nu] =  0 \\
 &T_{\mu \nu}[e,\omega] \equiv \partial_\mu e_\nu - \partial_\nu e_\mu  + [\omega_\mu, e_\nu] = 0 \;.
\end{cases}
\end{eqnarray}
Among these six classical equations, only two involve time derivatives and therefore can be interpreted as equations
of motion. As we will see in the following, the four remaining equations act as  first class constraints 
in the Hamiltonian framework and generate the gauge symmetries of the theory. 

\subsubsection{Symmetries: gauge symmetries and diffeomorphisms}

As the Chern-Simons formulation of three-dimensional gravity is a gauge theory
with local symmetry group $IG$, its action admits an infinite dimensional symmetry group 
${\cal G}=\cif({\cal M},IG)$  which acts on the connections according to
\be\label{gauge symmetry}
\forall \, g \in {\cal G} \;,\;\;\;\;\;\;\; A \; \mapsto \; A^g  = gAg^\inv + gdg^\inv \;.
\ee
The invariance of the action  $S_{SC}$  \eqref{CS action} with respect to these transformations
is an immediate consequence of the $Ad$- invariance of the bilinear form $\langle,\rangle$. It has been shown in \cite{Witten1} that they correspond to the infinitesimal diffeomorphism symmetries of gravity. This is most easily seen by 
rewriting the infinitesimal transformation laws (\ref{gauge symmetry}) in terms of the triad and spin connection 
\begin{eqnarray}
\label{infsym}
\delta e_\mu \; = \; \partial_\mu \baa \, + \, [\omega_\mu,\baa] \, + \, [e_\mu,\upsilon] \;\;\;\text{and} \;\;\;
\delta \omega_\mu \; = \; \partial_\mu \upsilon \, + \, [\omega_\mu,\upsilon]
\end{eqnarray}
where $g^\inv=(\upsilon,\baa) \in \cif({\cal M}, {\mathfrak g} \ltimes \mathbb R^3)$.  Setting  $\baa=\xi^\mu e_\mu$ and $\upsilon=\xi^\mu\omega_\mu$, one can then express these 
 transformations in terms of the Lie derivatives ${\cal L}_\xi$ along the vector field $\xi=\xi^\mu \partial_\mu$:
\begin{eqnarray}\label{infinitesimal symmetries}
\delta e_\mu \; = \; {\cal L}_\xi e_\mu \, + \, \xi^\nu T_{\mu \nu}[e,\omega] \;\;\;\;\text{and} \;\;\; 
\delta \omega_\mu \; = \; {\cal L}_\xi \omega_\mu \, + \, \xi^\nu F_{\mu \nu}[\omega] , 
\end{eqnarray}
where $F_{\mu\nu}[\omega]$ and $T_{\mu\nu}[e, \omega]$ are the curvature and torsion \eqref{flatcond} which vanish on the space of classical solutions.  This establishes the on-shell equivalence of infinitesimal  diffeomorphisms and infinitesimal Chern-Simons gauge transformations. Note, however, that this equivalence applies only to gauge transformations and diffeomorphisms which are connected to the identity, whereas the status of large (i.~e.~not infinitesimally generated) diffeomorphisms and gauge transformations is more subtle \cite{giu1,giu2, we1}.

\subsubsection{Canonical analysis}

On manifolds of topology  ${\cal M}=S \times \mathbb I$ one can give a Hamiltonian formulation of the theory.  For simplicity, we focus on the case where $S$ is an oriented two-surface of general genus. The case of a surface with punctures representing massive, spinning particles is a straightforward generalisation which is discussed extensively in the literature (see \cite{Carlipbook} and references therein). 

Decomposing the gauge field 
$A=A_tdt+A_i dx^i$
into a time component $A_t$ and a gauge field $A_S=A_idx^i$ on the spatial surface, 
we can rewrite 
the action (\ref{first order action})  as 
\begin{eqnarray}\label{CS Hamiltonian}
S_{CS}[A]= \int_{\mathbb I} dt \int_S d^2x \, \epsilon^{ij} \, 
\left( - \langle A_i, \, \partial_t A_j \rangle \, + \, \langle A_t,\, F[A]_{ij} \rangle \right) 
\end{eqnarray}
where $\epsilon^{ij}=\epsilon^{tij}$.
This implies that the phase space variables are the components of the spatial gauge field $A_S=A_idx^i$ and that their canonical Poisson brackets are given by
\begin{eqnarray}\label{Poisson bracket}
\{A^i_{\alpha}(x) , A^j_{\beta}(y)\}\,  \; = \; \,\epsilon^{ij} \, \delta^{(2)}(x-y) \,
\langle \xi_\alpha ,\xi_\beta \rangle,
\end{eqnarray}
where $\xi_\alpha \in \{J_a,P_b\}_{a,b=0,1,2}$ are the generators of the Lie algebra $\mathfrak g \ltimes \mathbb R^3$ and
$\delta^{(2)}(x-y)$ is the  delta distribution on $S$.  The time components 
$A_t$ of the gauge field act as Lagrange multipliers
which impose the six primary constraints ${\cal F}^\alpha(x) \equiv \epsilon^{ij}F_{ij}^\alpha[A(x)]=0$. 
It is easy to check that these primary constraints are first class and that the system 
admits no more constraints. They form a Poisson algebra, and they generate infinitesimal gauge symmetries.

When expressed in terms of the BF variables $e$ and $\omega$, the only non-trivial 
Poisson brackets in \eqref{Poisson bracket} are the ones which pair the components of the triad and spin connection
\begin{eqnarray}\label{BF Poisson bracket}
\{e_i^a(x), \omega_j^b(y)\} \; = \;  \eta^{ab} \, \epsilon_{ij} \, \delta^{(2)}(x-y).
\end{eqnarray}
Roughly speaking, the triad $e$ and the connection $\omega$ are canonically conjugated variables.
Moreover, by considering this expression, one finds that  the first class constraints can be grouped into the two sets
\begin{eqnarray}\label{continuous constraints}
F(x) \equiv \epsilon^{jk}F_{jk}[\omega(x)] \, = \, 0 \;\;\;\text{and} \;\;\; 
T(x) \, \equiv \, \epsilon^{jk} T_{jk}[e(x),\omega(x)] \, = \, 0 
\end{eqnarray}
which   generate the infinitesimal gauge symmetries given by (\ref{infsym}), \eqref{infinitesimal symmetries}
\begin{eqnarray}
\{\baa_a F^a(x) + \upsilon_a T^a(x),e_\mu(y)\}  & =  & \delta^{(2)}(x-y) \,\delta e_\mu(x) \\
\{\baa_a F^a(x) + \upsilon_a T^a(x),\omega_\mu(y)\} & = &  \delta^{(2)}(x-y) \,\delta \omega_\mu(x).\nonumber
\end{eqnarray}

{\bf The physical phase space}

To give a simple presentation of the physical phase space, it is advantageous to work with the Chern-Simons formulation of the theory. Let us recall that solutions of the constraints form an infinite dimensional affine
space,  the space of flat $IG$-connections on $S$  denoted  by ${\cal F}(IG,S)$. This space
inherits a Poisson bracket (\ref{Poisson bracket}) from the Chern-Simons action and the gauge symmetry action 
(\ref{gauge symmetry}). 
The physical phase space, denoted ${\cal P}(IG,S)$, is the moduli space of flat $IG$-connections modulo gauge transformations on the spatial surface $S$:
\begin{eqnarray}
{\cal P}(IG,S) \; \equiv \; {\cal F}(IG,S)/{\cal G}_S\qquad {\cal G}_S=\cif(S,IG)\;.
\end{eqnarray}
It inherits a symplectic structure from
the Poisson bracket on ${\cal F}(IG,S)$ and, remarkably, is  of finite dimension. 
More specifically, the physical phase space ${\cal P}(IG,S)$ can be parametrised by the holonomies along curves on the spatial surface $S$ and is isomorphic to the space 
$\text{Hom}(\pi_1(S),IG)/ IG$, where the quotient is taken with respect to the action of $IG$ by simultaneous conjugation.
The physical observables are, by definition, functions on ${\cal P}(IG,S)$. A basis can be constructed using the
notion of spin-networks on $S$. Alternatively, one can work with conjugation invariant functions of the holonomies along a set of curves on $S$ representing the elements of its fundamental group $\pi_1(S)$.
The Poisson bracket between two such observables was first described by Goldman \cite{Goldsymp}.

\section{Discretisation of the phase space}
\label{discretesect}

\subsection{Discretisation via graphs}

We are now ready to discuss the discrete descriptions of the phase space underlying three-dimensional loop quantum gravity and the combinatorial quantisation formalism, the latter of which is due to Fock and Rosly \cite{FR}. In both cases, the phase space is discretised by means of graphs embedded into the spatial two-surface, and the resulting descriptions are equivalent. However, as we will show in the following, there are important conceptual differences between the two discretisations which directly manifest themselves in the corresponding  quantisation approaches. 

\begin{figure}[h]
\psfrag{l1}{$\lambda_1$}
\psfrag{l2}{$\lambda_2$}
\psfrag{l3}{$\lambda_3$}
\psfrag{l4}{$\lambda_4$}
\psfrag{v}{$v$}
\centering
\includegraphics[scale=0.8]{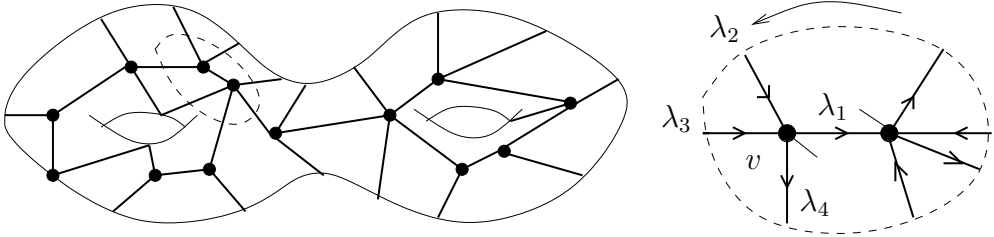}
\caption{\small{Illustration of the discretisation of a genus two surface $S$ by a graph $\Gamma$. 
On the right,
we focus on a particular part of $\Gamma$ where the structures of the graph have been highlighted: the edges are oriented
and the vertices are endowed with a cilium (the short thin lines) which defines a linear ordering of the incident edges. At the vertex $v$, we have 
we have $S(v)=\{\lambda_1,\lambda_4\}$ and $T(v)=\{\lambda_2,\lambda_3\}$;
$O(\lambda_1,s)<O(\lambda_2,t)<O(\lambda_3,s)<O(\lambda_4,t)$. }}
\label{graph}
\end{figure}

\label{graphsect}
We start by introducing the graphs used in the discretisation.
In the following we consider an oriented two-surface $S$ of general genus and with a general number of punctures together with an oriented graph $\Gamma$  embedded into the surface. We do not restrict attention to graphs associated with or dual to triangulations, but require that the graph is sufficiently refined to resolve the surface's topology.  We denote by $V_\Gamma$ and $E_\Gamma$ respectively the set of its vertices and the set of its oriented edges. 
For a given edge $\lambda\in E_\Gamma$ we denote by $s(\lambda)$ its starting vertex and by $t(\lambda)$ its target vertex and write $-\lambda$ for the edge with the opposite orientation.
For each vertex $v$, we introduce the set $S(v)=\{\lambda\in E_\Gamma\:| \; s(\lambda)=v\}$  of edges starting at $v$ and  the set 
$T(v)=\{\lambda\in E_\Gamma\:| \; t(\lambda)=v\}$ of edges ending at $v$,
as shown in Fig. \ref{graph}.

Such a graph is sufficient to define spin network functions and to formulate the three-dimensional version of loop quantum gravity. However,  for the combinatorial quantisation, additional structures are required. 
More precisely, we need a {\em ciliated fat graph}, which is obtained by adding a cilium at each vertex
of the oriented graph as shown in Fig.~\ref{graph}. As the orientation of the surface $S$ induces a cyclic ordering of the edges starting or ending in each vertex, the addition of the cilium defines a linear ordering of these edges. 

\begin{figure}[h]
\psfrag{l1}{$\lambda_1$}
\psfrag{l2}{$\lambda_2$}
\psfrag{l3}{$\lambda_3$}
\psfrag{l4}{$\lambda_4$}
\psfrag{l5}{$\lambda_5$}
\psfrag{l6}{$\lambda_6$}
\psfrag{l7}{$\lambda_7$}
\psfrag{l8}{$\lambda_8$}
\centering
\includegraphics[scale=0.6]{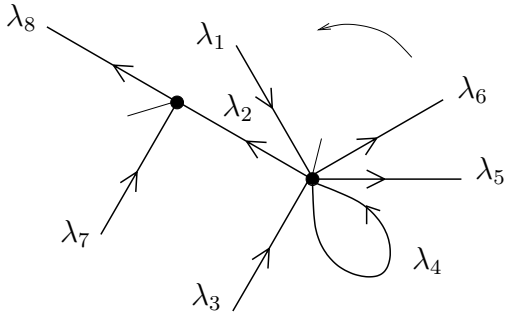}
\caption{\small{Illustrations of the sets $S^\pm$ and $T^\pm$ defined in (\ref{endedgedef}): 
$S^+(t(\lambda_1))=\{\lambda_2,\lambda_4,\lambda_5,\lambda_6\}$, $T^+(t(\lambda_1))=\{\lambda_3,\lambda_4\}$,
$T^-(s(\lambda_2))=\{\lambda_1\}$, $T^-(t(\lambda_2))=\{\lambda_7\}$}}
\label{sets}
\end{figure}

In the following we write $O(\lambda,s)<O(\tau,s)$ ($O(\lambda, s)<O(\tau, t)$) if $\lambda$ is an edge starting  at $v$ and of lower order than another edge $\tau$ starting (ending) at the same vertex and, analogously  $O(\lambda,t)<O(\tau,s)$ ($O(\lambda, t)<O(\tau, t)$) for edges  $\lambda$ that  end at the vertex, as shown in Fig.~\ref{graph}. 
We denote by $S^+(s(\lambda))$, $S^-(s(\lambda))$, respectively, the set of edges starting at the starting vertex of $\lambda$ and of higher and lower order than $\lambda$ and by $T^+(s(\lambda))$, $T^-(s(\lambda))$ the set of edges ending at the starting vertex of $\lambda$ and of higher and lower order than $\lambda$
\begin{align}
\label{endedgedef}
&S^+(s(\lambda))\!=\!\{ \eta\in S(s(\lambda))\!\!: \!O(\lambda,s)\!<\!O(\eta,s)\}\nonumber
&S^-(s(\lambda))\!=\!\{ \eta\in S(s(\lambda))\!\!: \!O(\lambda,s)\!>\!O(\eta,s)\}\\
&T^+(s(\lambda))\!=\!\{ \eta\in T(s(\lambda))\!\!: \!O(\lambda,s)\!<\!O(\eta,t)\}
&T^-(s(\lambda))\!=\!\{ \eta\in T(s(\lambda))\!\!: \!O(\lambda,s)\!>\!O(\eta,t)\}\nonumber
\end{align}
Analogously, we define the sets $S^\pm(t(\lambda))$, $T^\pm(t(\lambda))$. Note that these definitions are also valid for edges $\lambda, \eta\in E_\Gamma$ that are loops based at a vertex of the graph. For instance,  the set
\begin{align}
S^+(s(\lambda))\cap T^+(s(\lambda))=\{\eta\in S(s(\lambda))\cap T(s(\lambda))\,|\, O(\eta,s), O(\eta,t)>O(\lambda,s)\}
\end{align}
denotes the set of loops $\eta$ based at the starting vertex of $\lambda$ for which both ends are of higher order than $\lambda$.
If $\lambda$ is a loop, we write 
\begin{align}
S^+(s(\lambda))\cap S^-(t(\lambda))=\{\eta\in S(s(\lambda)\,|\, O(\lambda,s)<O(\eta,s)<O(\lambda,t)\}
\end{align}
for the set of edges that lie between the two ends of $\lambda$ with respect to the ordering at the vertex $s(\lambda)=t(\lambda)$. These sets are illustrated in Fig. \ref{sets}.

\subsection{Phase space variables}

In the discrete description of the phase space, the continuous dynamical variables,  the connection $A(x)$ in the Chern-Simons formulation
and the triad $e(x)$ and the spin-connection $\omega(x)$ in the BF formulation of the theory, are replaced by ``non-local" 
variables associated to oriented paths on the spatial surface $S$. 
In three-dimensional loop quantum gravity, these variables are obtained  by integrating the 
$G$-connection $\omega$ and 
the triad $e$ over general paths $\gamma:[0,1]\rightarrow S$ on the spatial surface. 
This amounts to assigning a group element $u_\gamma \in G$ and a vector $\bq_\gamma\in \RR^3$ to each 
path $\gamma$ 
\begin{align}
\label{uqdef}
&u_\gamma=\text{Pexp}\int_{\gamma} \omega_\mu \, dx^\mu\qquad \text{and} 
\qquad q^a_\gamma=\int_{\gamma} e^a_\mu \, dx^\mu \,.
\end{align}
In the Chern-Simons formulation, triad and spin connection are combined into a Chern-Simons gauge field. This makes it natural to work with $IG$-valued phase space variables obtained by integrating the Chern-Simons gauge field $A$ along paths on $S$. Parametrising elements of the three-dimensional Euclidean and Poincar\'e groups as in \eqref{gparam}, one assigns a $G$-element $u_\gamma$ and a vector $\bj_\gamma\in\RR^3$ to each path $\gamma$ 
\begin{align}
\label{csvars}
H_\gamma=(u_\gamma, -\Ad(u_\gamma)\bj_\gamma)=\text{Pexp}\int_\gamma A_\mu \, dx^\mu.
\end{align}
The variables obtained by reversing the orientation of the path $\gamma$ are then related to the original variables as follows
\begin{align}
\label{frinvert}
u_{-\gamma}=u_\gamma^\inv\, ,\qquad \bj_{-\gamma}=-\Ad(u_\gamma)\bj_\gamma,
\qquad \bq_{-\gamma}=-\bq_{\gamma}\,.
\end{align}
From the definition of the gauge field $A$, it is easy to see that the $G$-valued variables $u_\gamma$ agree with the ones used in loop quantum gravity and defined in \eqref{uqdef}.
Moreover, a short calculation shows that the vectors $\bj_\gamma$ are given in terms of the triad and the spin-connection by
the relation
\begin{align}
\label{jwedef}
\bj_\gamma=\int_\gamma \Ad(u_\gamma^\inv(y)) \, e_\mu(y) \, dy^\mu,
\end{align}
where $u_\gamma(y)$  denotes the path ordered exponential along $\gamma$ from the starting point $s(\gamma)$ 
to $y \in \gamma$. We see that there is a priori no simple and explicit relation between the vectors $\bj_\gamma$ 
and $\bq_\gamma$ at the classical level.
However, we will demonstrate in Section~\ref{jqrel} that the associated operators on the Hilbert spaces of the quantum theory exhibit a direct and physically intuitive relation.

\subsection{Poisson structure}

In the description of the phase space underlying the loop quantum formalism, the canonical Poisson structure \eqref{BF Poisson bracket}  induces a bracket on functions of the group elements $u_\gamma$
and the vectors $\bq_\tau$ associated to paths $\gamma,\tau:[0,1]\rightarrow S$  which intersect transversally in a vertex.
 From the canonical Poisson bracket (\ref{BF Poisson bracket}) of the triad and spin-connection, 
it follows that the bracket of functions $f_\gamma, g_\tau$ of the $G$-elements $u_\gamma$, $u_\tau$ 
 vanishes
 \begin{align}
&\{f_\gamma,g_\tau\}=0.
\end{align}
Similarly, one has for the bracket of the associated vectors $\bq_\gamma,\bq_\tau$
\begin{align}
\{q_\gamma^a,q_\tau^b\}=0.
\end{align}
The only non-trivial brackets are those of functions of the $G$-elements  $u_\gamma$ with vectors $\bq_\tau$.
A standard calculation, see for instance \cite{Carlipbook}, yields
\begin{align}
\label{pathbrack}
\{q^a_\gamma, f\}(u_{\tau_2}u_{\tau_1})=\frac{d}{dt}|_{t=0} f(u_{\tau_1}e^{tJ_a} u_{\tau_2}),
\end{align}
where  $\tau=\tau_2\circ \tau_1$ and  $t(\tau_1)=s(\tau_2)$
is the intersection point between $\tau$ and $\gamma$.
Note that this bracket is only defined for paths $\gamma,\tau$ which intersect transversally, i.~e.~for which the oriented intersection number is well-defined. 

In the combinatorial formalism, the issue of the Poisson structure is more subtle.
This is partly due to the fact that one works with $IG$-valued holonomy variables, which combine 
the $G$-holonomies $u_\lambda\in G$ and the vectors $\bj_\lambda\in\RR^3$ and whose brackets are  intrinsically more complicated. Moreover,  one cannot restrict attention to transversally intersecting paths but also needs to consider paths which meet in their starting and end points.
Expanding the path ordered exponential \eqref{csvars} does not yield a well-defined expression for the Poisson bracket of such variables due to the presence of delta-distributions at the end points. This implies that the canonical Poisson structure associated to the action does not induce a Poisson structure of these variables. 

A regularisation of these ill-defined Poisson brackets is provided by the formalism of Fock and Rosly \cite{FR}. This regularisation requires a graph $\Gamma$ endowed with a ciliation which induces a linear ordering of the edges incident at each vertex of $\Gamma$ as defined in Subsection \ref{graphsect}. The other central ingredient is a classical $r$-matrix for the gauge group $IG$, which is explained in appendix \ref{frphsp}.
It has been shown by Fock and Rosly \cite{FR} that together with the ciliation such a classical $r$-matrix 
allows one to define a consistent Poisson bracket on the variables obtained by integrating the Chern-Simons gauge field along the edges of the graph and that this auxiliary Poisson structure induces the canonical Poisson structure on the physical phase space. A summary of Fock and Rosly's Poisson structure \cite{FR} and its application to three-dimensional gravity is given in appendix \ref{frphsp}.

When applying Fock and Rosly's Poisson structure to three-dimensional gravity, one finds 
 the Poisson bracket of the $G$-holonomies associated to 
different paths on $S$ vanish as they do in the loop formalism. More generally, we have
\begin{align}
\label{FRbrack1}
\{f,g\}=0\qquad\forall f,g\in\cif(G^{|E_\Gamma|}),
\end{align}
where the arguments of $f$ and $g$ are identified with the $G$-holonomies $u_\lambda$ along the edges $\lambda\in E_\Gamma$.  
The bracket of vectors $\bj_\lambda$ with functions $f\in\cif(G^{|E_\Gamma|})$ is given by certain vector fields 
$\bX_\lambda$ on the manifold $G^{|E_\Gamma|}$ which will be described explicitly below 
\begin{align}
\label{jfbrack0}
\{j^a_\lambda, f\}=X^a_\lambda f\qquad\forall f\in\cif(G^{|E_\Gamma|}).
\end{align}
The brackets between the vectors $\bj_\lambda$ are given by the Lie bracket of the associated vector fields and can be determined explicitly via the Jacobi identity 
\begin{align}
\label{FRbrack2}
\{\{j_\lambda^a, j_\tau^b\}, f\} =\{j^a_\lambda,\{j^b_\tau, f\}\}-\{j^b_\tau,\{j_\lambda^a, f\}\}=(X^a_\lambda X^b_\tau-X^b_\tau X^a_\lambda)f=[X^a_\lambda, X^b_\tau]f.
\end{align}

\begin{figure}[h]
\psfrag{c1}{(a)}
\psfrag{c2}{(b)}
\psfrag{c3}{(c)}
\psfrag{c4}{(d)}
\psfrag{l}{$\lambda$}
\psfrag{t}{$\tau$}
\centering
\includegraphics[scale=0.6]{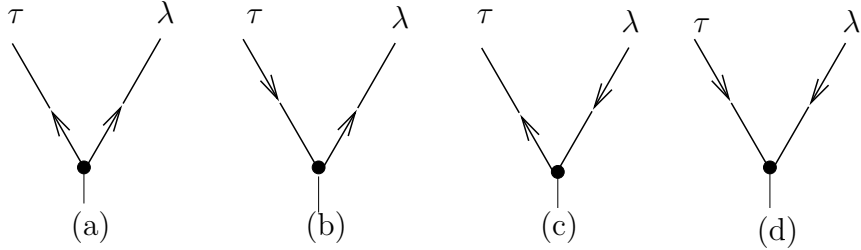}
\caption{\small{The four different configurations for two edges meeting at a vertex.}}
\label{edgecase}
\end{figure}

In order to give explicit expressions for the  vector fields $X^a_\lambda$, we need to introduce some notations. In the following, we write 
$f_\lambda\in\cif(G^{|E_\Gamma|})$  for a function that depends only on the group element $u_\lambda$ associated to a given edge $\lambda\in E_\Gamma$. We denote by
$L_\lambda^a$ and $R_\lambda^a$, respectively, the right- and left-invariant vector fields \eqref{vecfieldact} corresponding to the variable
$u_\lambda$
\begin{align}
\label{compvecfields}
&R_\lambda^af_\tau = R^a f_\tau \,\,\, \text{if $\tau=\lambda$, otherwise} \;\;\; R_\lambda^af_\tau = 0\\
&L_\lambda^af_\tau = L^a f_\tau \,\,\, \text{if $\tau=\lambda$, otherwise} \;\;\; L_\lambda^af_\tau = 0\nonumber.
\end{align}
By applying Fock and Rosly's prescription to the case at hand,  we then obtain expression for the Poisson brackets and the vector fields $X^a_\lambda$ 
 \eqref{jfbrack0} in terms of  these right- and left-invariant vector fields 
\begin{align}
\label{jbrack}
\{j^a_\lambda, f\}=X^a_\lambda f =  -R_\lambda^a f-\!\!\!\!\!\!\!\!\!\sum_{\tau\in S^+(s(\lambda))}\! \!\!\!\!\!\!R_\tau^af -\!\!\!\!\!\!\sum_{\tau\in T^+(s(\lambda))} \!\!\!\!\!\!\!L_\tau^af 
+\Ad(u_\lambda^\inv)^{a}_{\;\;b}\left(\sum_{\tau\in S^+(t(\lambda))} \!\!\!\!\!\!\!R_\tau^bf +\!\!\!\!\!\!\sum_{\tau\in T^+(t(\lambda))} \!\!\!\!\!\!\!L_\tau^bf \right).
\end{align}
Although the general formula is rather complicated,
the action of the vector fields $X^a_\lambda$ on  the group elements $u_\tau$, $\tau \in E_\Gamma$, corresponds to a simple and intuitive geometrical prescription:
\begin{enumerate}
\item Group elements $u_\tau$ associated to edges $\tau$ which do not have a vertex in common with 
$\lambda$ are unaffected by the action of $X^a_\lambda$.
\item Group elements $u_\tau$ associated to edges $\tau$ which do have a vertex in common with $\lambda$ but are of lower order at this vertex are unaffected.
\item $X^a_\lambda$ acts on the group element $u_\lambda$ by right multiplication
$X^a_\lambda f_\lambda=-R^af_\lambda$.
\item $X^a_\lambda$ acts  on the group elements $u_\tau$ associated with edges $\tau\in S^+(s(\lambda))$ which start at the starting vertex $s(\lambda)$ and are of higher order than $\lambda$ (case $a$ in Fig.\ref{edgecase})
by right multiplication: $X^a_\lambda f_\tau =-R^af_\tau$.
\end{enumerate}

These rules allow one to compute the action of the vector fields $X^a_\lambda$ on any function 
$f\in\cif(G^{|E_\Gamma|})$.
In particular, its action on edges that end at the starting vertex of $\lambda$ or start or end at its 
target vertex (cases $b$, $c$, $d$ in Fig.~\ref{edgecase}, respectively) is obtained by using formula 
\eqref{frinvert} to invert the orientation of the edges. This yields
\begin{align}
& X^a_\lambda f_\tau=-L^af_\tau & &\tau\in T^+(s(\lambda))\;\text{(case b)}\\
&X^a_\lambda f_\tau=\Ad(u_\lambda^\inv)^a_{\;\;b} R^b f_\tau  & &\tau\in S^+(t(\lambda))\;\text{(case c)}\\
& X^a_\lambda f_\tau=\Ad(u_\lambda^\inv)^a_{\;\;b} L^bf_\tau & &\tau\in T^+(t(\lambda)) \;\text{(case d)}.
\end{align}

Note that this prescription is also defined for loops that start and end at the same vertex or for loops that have two vertices in common. In this case, one simply applies the prescription above to both ends of the edges and adds the resulting expressions.

\begin{figure}[h]
\psfrag{k}{$\kappa$}
\psfrag{g}{$\gamma$}
\psfrag{t}{$\tau$}
\centering
\includegraphics[scale=0.6]{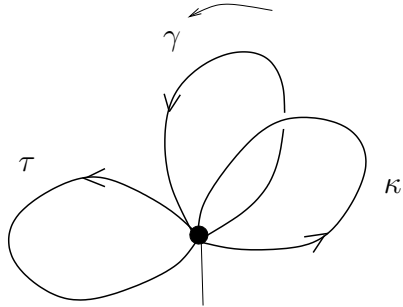}
\caption{\small{Examples of  a ciliated graphs with edges that are loops. }}
\label{loopex}
\end{figure}

\begin{example}
As an example,  we consider the configuration with three loops $\gamma,\kappa,\tau$ represented in Fig. \ref{loopex}.  The linear ordering is such that $O(\kappa,s)<O(\gamma,s)<O(\kappa,t)<O(\gamma,t)<O(\tau,s)<O(\tau,t)$. Applying  formula \eqref{jbrack}, one finds that the
Poisson brackets between the associated loop variables $\bj_\gamma,\bj_\kappa,\bj_\tau$ with functions of the holonomies $u_\gamma, u_\kappa, u_\tau$ are given by
\begin{align}
\label{ex2brack}
&\{j^a_\kappa, f_\kappa\}=-(R^a+L^a)f_\kappa\\
\text{Action of $X_\kappa$:} \;\;
 &\{j^a_\kappa, f_\gamma\}=- \left((\delta^a_b-\Ad(u_\kappa^\inv))^a_{\;\;b}L^b +R^a\right)f_\gamma\nonumber\\
&\{j^a_\kappa, f_\tau\}=-( (\delta^a_b-\Ad(u_\kappa^\inv))^a_{\;\;b}(L^b+R^b)f_\tau\nonumber\\
\nonumber\\
&\{j^a_\gamma, f_\kappa\}=-L^af_\kappa\\
\text{Action of $X_\gamma$:} \;\; &\{j^a_\gamma, f_\gamma\}=-(R^a+L^a)f_\gamma\nonumber\\
 &\{j^a_\gamma, f_\tau\}=- (\delta^a_b-\Ad(u_\kappa^\inv))^a_{\;\;b}(L^b+R^b)f_\tau\nonumber\\
 \nonumber\\
 &\{j^a_\tau, f_\kappa\}=0\\
\text{Action of $X_\tau$:} \;\; &\{j^a_\tau,f_\gamma\}=0\nonumber\\
 &\{j^a_\tau, f_\tau\}=-(R^a+L^a)f_\tau\nonumber,
\end{align} 
where $L^a$, $R^a$ are the right- and left-invariant vector fields \eqref{vecfieldact} on $G$.
For  functions  $f_\kappa, f_\gamma, f_\tau\in\cif(G^{|E_\Gamma|})$ which are invariant under conjugation, i.~e.~physical observables,  the only non-vanishing brackets in \eqref{ex2brack} are 
\begin{align}
\label{ex22brack}
 &\{j^a_\kappa, f_\gamma\}=\Ad(u_\kappa^\inv)^a_{\;\;b}L^b f_\gamma & &\{j^a_\gamma, f_\kappa\}=-L^af_\kappa\end{align} 
This agrees with the result derived from formula \eqref{pathbrack} and demonstrates the dependence of the brackets on intersection points evident there. It is a manifestation of the fact that the Fock and Rosly bracket of graph gauge invariant functions
is identical to the canonical bracket on the physical phase space.
\end{example}

\subsection{Physical phase space} 

\label{physphspace}
We are now ready to discuss the implementation of the constraints and the construction of the physical phase space.
In both formalisms,  the construction of the physical Hilbert space requires the implementation of 
 a discrete version of the constraints \eqref{flatcond}. These are obtained by integrating \eqref{flatcond} along each closed, contractible loop $\gamma$ on the spatial surface $S$ and reflect the topological nature of the theory
\begin{align}
\label{loopconstdisc}
F[\gamma]=u_\gamma\approx 1\qquad  T^a[\gamma]=\Ad(u_\gamma)^{a}_{\;\;b} j^b_\gamma=\int_\gamma \Ad^{a}_{\;\;b}( u_\gamma u_\gamma^\inv(y)) e^b_\mu(y)  dy^\mu\approx 0.
\end{align}
The constraint $F[\gamma]$ corresponds to the flatness condition $F_{\mu\nu}[\omega]=0$ and the constraint $T^a$ to the Gauss constraint  $T_{\mu\nu}[e,\omega]=0$ in \eqref{flatcond}.
In the Chern-Simons formulation of the theory, these conditions are combined into the requirement that the $IG$-valued  holonomy $H_\gamma$ given by  \eqref{csvars} is trivial
for any  contractible loop $\gamma$ on $S$.

In the loop formalism, the construction of the  physical phase space is usually not discussed separately on the classical level but
follows from the corresponding discussion for the quantum theory. The general idea is to select certain paths $\gamma$ on the spatial surface $S$ which form a graph and to consider the associated discretised variables $u_\gamma$, $\bq_\gamma$ defined as in \eqref{uqdef}.  
While both the Gauss constraint $F[x]$ and the Hamiltonian constraints $T^a[x]$ are discretised by integrating them along loops on the spatial surface as in \eqref{loopconstdisc}, different paths are chosen for this discretisation: For the Gauss constraint $T^a[\gamma]$ one selects small closed loops $\gamma$ around the vertices of the graph which intersect its edges transversally.  The discrete version $F[\gamma]$ of the Hamiltonian constraint is obtained by integrating it  along  closed loops in the graph itself.
One then obtains a set of discrete constraints which generate discrete gauge transformations acting on the variables $u_\gamma,\bq_\gamma$.  The details then depend on the choice of the paths and the choice of the discretisation, and there appears to be no standard convention in the literature.
A detailed investigation of these gauge transformations and the construction of the physical Hilbert space for a particular choice of such a discretisation is given in \cite{LPR}.

In the combinatorial formulation, the situation is more involved, as one works with $IG$-valued holonomies associated to a fixed graph. To discuss the constraints and the construction of the physical phase space, one considers the space of graph connections  ${\cal A}_\Gamma \equiv \cif(G^{|E_\Gamma|}) \otimes {\cal J}_\Gamma$
with ${\cal J}_\Gamma=\{\bj_\lambda \vert \lambda \in E_\Gamma\}$, which consists of assignments of $IG$-valued holonomies $H_\tau$ to each edge $\tau\in\Gamma$ and can be viewed as a discrete version of 
the space of $IG$-connections on the surface $S$. Similarly,
the discrete version of the space of flat connections ${\cal F}_\Gamma$ is the space of flat graph 
connections and is obtained from the space of graph connections by imposing the constraint of vanishing $IG$-holonomy  
for all closed, {\em contractible} loops $\ell=\lambda_n\circ \ldots \circ \lambda_1$ of $\Gamma$
\begin{align}
\label{constgen}
\prod_{\lambda \in \ell} (u_\lambda, -\Ad(u_\lambda)\bj_\lambda)\approx 1,
\end{align}  where the product runs over
the edges $\lambda_n,\ldots,\lambda_1$ in the loop $\ell$ in the order in which they appear in the loop.
The $G$-component and the translational component of this constraint correspond to the variables \eqref{loopconstdisc} for $\gamma=\ell$ and are given by
\begin{eqnarray}\label{discrete constraints}
F_\ell =(u_{\lambda_n}\cdots u_{\lambda_1},0) \approx 1 \;\;\;\; \text{and} \;\;\;\;
T_\ell = (1,\sum_{i=1}^n\Ad((u_{\lambda_{i-1}}\cdots u_{\lambda_1})^\inv)\bj_{\lambda_i})\approx0.\end{eqnarray}
There is also a discrete version of the 
 group of gauge transformations ${\cal G}$: the group ${\cal G}_\Gamma$ of graph gauge transformations which is isomorphic to
$IG^{|V_\Gamma|}$.  A graph gauge transformation is an assignment of an $IG$-element
$G_v=(g_v,-\Ad(g_v)\bx_v)$ to each vertex $v\in V_\Gamma$. Its action on the graph connections is given by 
\begin{align}
\label{ggtrafo}
H_\lambda\mapsto G_{t(\lambda)}\cdot H_\lambda \cdot G_{s(\lambda)}^\inv
\end{align}
or, equivalently,
\begin{align}
&u_\lambda\mapsto g_{t(\lambda)}\cdot u_\lambda \cdot g^\inv_{s(\lambda)}\label{utrafo}\\
&\bj_\lambda\mapsto \Ad(g_{s(\lambda)})(\bj_\lambda-\bx_{s(\lambda)})+\Ad(g_{s(\lambda)} u_\lambda^\inv)\bx_{t(\lambda)}\label{jtrafo}.
\end{align}
For any sufficiently refined graph $\Gamma$, the phase space of the theory which is the moduli space
of flat $IG$-connections on the surface $S$ modulo gauge transformations is isomorphic to the quotient of the space ${\cal F}_\Gamma$ of flat graph connections modulo
graph gauge transformations:
\begin{eqnarray}
{\cal P}(IG,S) \; \simeq \; {\cal F}_\Gamma/{\cal G}_\Gamma \;.
\end{eqnarray}

The central result of Fock and Rosly \cite{FR} is that the Poisson structure given by equations \eqref{FRbrack1}, \eqref{jfbrack0}, \eqref{FRbrack2} descends to this quotient and induces the non-degenerate symplectic 
form on the moduli space of flat connections. In other words: physical observables are represented by functions on 
${\cal F}_\Gamma$ which are invariant under the graph gauge transformations ${\cal G}_\Gamma$, and the Poisson bracket of such observables agrees with the one given by the Fock-Rosly Poisson structure.  As a result, the symplectic form depends neither on the choice of
the (sufficiently refined) graph $\Gamma$, nor on the choice of the cilia on the vertices. 
In that sense, the description by Fock and Rosly \cite{FR} is an exact discretisation of Chern-Simons theory. Moreover, it can easily be extended to the case of surfaces with punctures representing massive point particles.  The only modification required is an additional set of constraints  similar to \eqref{constgen} which restrict the $IG$-holonomies of loops around particles to fixed  $IG$-conjugacy classes 
\begin{align}
\label{partconstgen}
\prod_{\lambda \in \ell} (u_\lambda, -\Ad(u_\lambda)\bj_\lambda)\in \mathcal C_{i},
\end{align} 
where $\ell$ is a loop around the $ith$ particle and $\mathcal C_i$ the $IG$-conjugacy class associated to this particle.

These results allow one to choose a minimal simplicial decomposition of $S$ for the graph $\Gamma$, i.~e.~ a set of generators of the fundamental group $\pi_1(S)$. 
This is the
starting point of the combinatorial quantisation of three dimensional gravity. However, as the purpose of this paper is a comparison between the combinatorial quantisation and  loop quantum gravity, the latter of which is based on the space of cylindrical functions on general graphs,  we will not restrict attention to such graphs in the following. A detailed discussion of the relation between general ciliated graphs and minimal simplicial decompositions is given in Sect.~\ref{physhilb}.

\section{Hilbert spaces and operators}

\label{hilbsect}

\subsection{Quantum states and kinematical Hilbert spaces}

\label{kinhilb}

In both formalisms, the quantisation proceeds in two steps.  
The first is to promote the discrete graph variables to an algebra of operators and to determine its 
unitary irreducible representations, which define the space of quantum states.  In both cases, 
the quantum states form the so-called space of cylindrical functions on $\Gamma$ which is the space $\cif(G^{|E_\Gamma|})$ of functions of the $G$-valued holonomies assigned to the edges of the graph. 
Note that the topological nature of the theory in three dimensions allows one 
to restrict attention to a single graph as long as it is sufficiently refined to resolve the topology of $S$.
The resulting quantum theory will be independent of the choice of the graph.

The second step is the construction of the kinematical and physical Hilbert spaces. This is done by promoting the constraints to operators acting on the space of cylindrical functions  $\cif(G^{|E_\Gamma|})$. Schematically,  kinematical states are the kernel of the quantum operators associated to the discretised version of the torsion $T(x)$.
Physical states are kinematical states which are in the kernel of the operators corresponding to the curvature $F(x)$
(\ref{continuous constraints}). 

In this Section, we focus on the space of quantum states and the construction of the kinematical Hilbert spaces in both approaches. We relate the fundamental quantum operators acting on these spaces and show how this relation provides a clear physical interpretation of the operators in the combinatorial formalism from the viewpoint of loop quantum gravity.
The construction of the physical Hilbert space  is discussed in Sect.~\ref{physhilb}.

\subsubsection{Loop quantum gravity} 

\begin{figure}[h]
\psfrag{l1}{$\lambda_1$}
\psfrag{l2}{$\lambda_2$}
\psfrag{l3}{$\lambda_3$}
\psfrag{l4}{$\lambda_4$}
\psfrag{l2p}{$\lambda_2'$}
\centering
\includegraphics[scale=0.8]{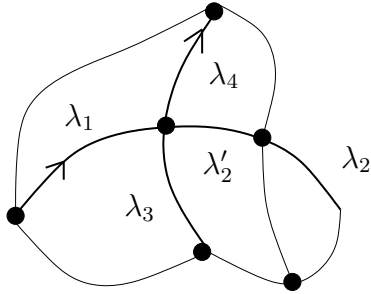}
\caption{\small{Illustration of a case where the derivative operator $\bq_\gamma$ has
a non-trivial action on a quantum state whose support is a graph $\gamma'$: $\gamma=\lambda_2\lambda_2'\lambda_1$
and $\gamma'=\lambda_4\lambda_3$. The operator $\bq_\gamma$ acts schematically on the common vertex $\gamma \cap \gamma'$.}}
\label{trans}
\end{figure}

In loop quantum gravity,  a quantum state is a priori any function
of the spin-connection $\omega$, and the two basic operators are  the spin connection $\omega$ and the triad $e$. The former acts by multiplication and the latter as a derivative operator
\be
e^i_a(x) \; = \; -i \epsilon_{ab} \, \eta^{ij} \, \frac{\delta}{\delta \omega_b^j(x)}.
\ee
However, many arguments \cite{ALreview,carlobook,thomasbook} lead to the conclusion that a quantum state is in fact a function of the $G$- valued holonomies obtained by integrating $\omega$ along the edges of the graph. The space of quantum states is thus the space $\cif(G^{|E_\Gamma|})$ of cylindrical functions for $\Gamma$
 endowed with the $L^2(G^{|E_\Gamma|})$ norm
\begin{align}
\label{physscalprod}
\langle \psi, \phi\rangle = \int d\mu(u_1)\cdots d\mu(u_{|E_\Gamma|}) \;
\overline{\psi(u_1,\ldots, u_{|E_\Gamma|})} \phi(u_1,\ldots, u_{|E_\Gamma|})
\end{align}
where $d\mu$ is the Haar measure on $G$.  
The basic discrete variables of loop quantum gravity (\ref{uqdef}) are cylindrical functions associated with $\Gamma$ and the quantum counterparts of the variables $\bq_\gamma$ in \eqref{uqdef}. The former act by multiplication, which can easily seen to be unitary with respect to the norm \eqref{physscalprod}
\be\label{multi}
\Pi(F)\psi \; = \; F\cdot \psi \;.
\ee
The action of the operators $\bq_\gamma$ is more subtle: As in the classical theory,  the action of $\bq_\gamma$ on a variable $u_{\gamma'}$ is well-defined if and only if 
the paths $\gamma$ and $\gamma'$ admit a well-defined intersection number, i.~e.~they cross transversally.
Thus, the action of $\bq_\gamma$ is not well-defined when $\gamma$ is a single edge of $\Gamma$;  the path $\gamma$ has to be the composition of at least
two edges. For instance, the action of $\bq_{\lambda_2\lambda_2'\lambda_1}$ on 
a state $\psi(u_{\lambda_4}u_{\lambda_3})$
where $t(\lambda_1)=s(\lambda_2)=t(\lambda_3)=s(\lambda_4)$, as illustrated in Fig. \ref{trans}, is given by
\be \label{deriv}
\Pi(q^a_{\lambda_2 \lambda_2'\lambda_1}) \psi\,(u_{\lambda_3}u_{\lambda_4}) \; = \; 
i\frac{d}{dt}|_{t=0} \psi(u_{\lambda_3}e^{tJ_a} u_{\lambda_4}) \;.
\ee
This formula is a direct quantisation of the Poisson bracket (\ref{pathbrack}). Its extension
to general paths is immediate, and it follows that the operators $\bq_\gamma$ act as vector fields
on the space of cylindrical functions. Together, (\ref{multi}) and (\ref{deriv}) provide an unitary representation $\Pi$ of the algebra of quantum operators
on the space of cylindrical functions on the graph $\Gamma$.

The kinematical Hilbert space $H_{kin}$ is obtained as the set of solutions of the quantum Gauss constraint and its construction is  well-understood. Kinematical states are  functions $\psi\in\cif(G^{|E_\Gamma|})$ of the $G$-holonomies along the edges of $\Gamma$ that satisfy the invariance condition
\begin{eqnarray}\label{kin invariance}
\psi(u_{\lambda_1},...,u_{\lambda_{|E_\Gamma|}}) =  
\psi(g_{s(\lambda_1)}^{-1} u_{\lambda_1} g_{t(\lambda_1)},...,g_{s(\lambda_{|E_\Gamma|})}^{-1} u_{\lambda_{|E_\Gamma|}} g_{t(\lambda_{|E_\Gamma|})})
\;\forall  g=(g_{v_1},..., g_{|V_\Gamma|}) \in G^{\vert V_\Gamma \vert} \!\!.
\end{eqnarray}
Due to left and right invariance of the Haar measure on $G$, the norm
(\ref{physscalprod}) is compatible with the quotient and induces a norm on $H_{kin}$.
In the case $G=SU(2)$, a dense basis of $H_{kin}$ is provided by the spin network functions. Spin network functions are constructed  by assigning a representation of $G$ to each edge $e\in E_\Gamma$ and an intertwiner to each vertex 
$v\in V_\Gamma$. In the case $G=SU(1,1)\cong SL(2,\mathbb R)$, the situation is more involved due to the non-compactness of the group.  Firstly, finite-dimensional irreducible representations of $SL(2,\mathbb R)$
are never unitary unless they are trivial. Instead, there are several series of
infinite-dimensional irreducible unitary representations labelled by continuous
parameters $\mu\in\mathbb R$.
 Moreover, the Peter-Weyl theorem,  which implies for compact Lie groups $G$ that
 that the spin network functions are dense in $L^2(G^{|E_\Gamma|})$,
 does not hold. The definition of spin network functions therefore
has to be undertaken within the framework of harmonic analysis.
 The presence
of representations labelled by continuous parameters then raises issues
of convergence whenever sums over discrete representation labels in the compact
case are replaced by integrals over continuous parameters. Another
source of divergences are integrals over the group $SU(1, 1) \cong SL(2,R)$  such
as the ones arising in the definition of the inner product. The construction
of spin networks for this group has been investigated in  \cite{FL,AO}.

The representation $\Pi$ defined in \eqref{multi}, \eqref{deriv} gives  a representation of kinematical operators acting on $H_{kin}$.  An important kinematical operator is the quantum counterpart of the   classical length of a path $\gamma:[0,1]\rightarrow S$
\be
L_\gamma \; = \; \int_\gamma ds \, \sqrt{\vert \eta_{ab}\,e^a\, e^b \vert}
\ee
The standard quantisation \cite{FLR} is such that spin-network states $\psi_\Gamma$ are eigenstates of the associated operator. It has been found in  \cite{FLR} that
its spectrum is discrete in the Euclidean case while it has discrete (for timelike curves)
and continuous (for spacelike curves) sectors in the Lorentzian case.

\subsubsection{Combinatorial formalism}

In the combinatorial formalism, the particularly simple  structure of the classical Poisson algebra for vanishing cosmological constant allows one to construct the kinematical Hilbert space and kinematical operators in a straightforward way. This is  due to the fact that the Poisson brackets of functions $f\in\cif(G^{|E_\Gamma|})$ vanish while the vectors $\bj_\lambda$ are identified with certain vector fields acting on functions $f\in\cif(G^{|E_\Gamma|})$.  The classical Poisson algebra is therefore of the type considered in Sect.~3.1. in \cite{we2} and can be quantised via the formalism established there, see in particular Theorem 3.1, Theorem 3.3. and Theorem 3.4.

By applying these results, one finds that the space of quantum states is the same as in the loop formalism, the space $\cif(G^{|E_\Gamma|})$ of cylindrical functions associated to the graph $\Gamma$ equipped with the norm \eqref{physscalprod}.
The basic quantum operators are the cylindrical functions $F\in\cif(G^{|E_\Gamma|})$ which act by multiplication as in \eqref{multi}
and the quantum counterparts of the vectors $\bj_\lambda$, $\lambda\in E_\Gamma$,  whose action on 
the states is given by:
\begin{align}\label{jact}
\Pi(j^a_\lambda) \psi=&i\{j_\lambda^a,\psi\}\\
=&-iR_\lambda^a \psi-\!\!\!\!\!\!\!\!\!\sum_{\tau\in S^+(s(\lambda))} \!\!\!\!\!\!\!iR_\tau^a\psi -\!\!\!\!\!\!\!\!\sum_{\tau\in T^+(s(\lambda))} \!\!\!\!\!\!\!iL_\tau^a\psi+\Ad(u_\lambda^\inv)^{a}_{\;\;b}\left(\sum_{\tau\in S^+(t(\lambda))} \!\!\!\!\!\!\!iR_\tau^b\psi +\!\!\!\!\!\!\!\!\!\sum_{\tau\in T^+(t(\lambda))} \!\!\!\!\!\!\!iL_\tau^b\psi \!\right).\nonumber
\end{align}
In contrast to the situation in loop quantum gravity,  the representation $\Pi(\bj^a_\lambda)$ of these operators is well-defined when $\lambda$ is a single edge 
of the graph $\Gamma$.  

The kinematical Hilbert space $H_{kin}$ is obtained by imposing invariance under the graph gauge transformations \eqref{utrafo} and hence characterised by \eqref{kin invariance} as in the loop formalism.  
The basic kinematical operators are  functions $F\in\cif(G^{|E_\Gamma|})$ satisfying  \eqref{kin invariance},  which act by multiplication,  and operators $\bJ$ that are linear combinations of the variables $\bj_\lambda$ with cylindrical functions
 as coefficients and preserve \eqref{kin invariance}. The latter can be identified with the vector fields on $G^{|E_\Gamma|}$ whose flow commutes with the action of the constraints $T_\ell$. 

 Two fundamental kinematical operators are the ``mass" operator $m_\ell$ and ``spin" operator $s_\ell$  associated to  closed loops 
 $\ell=\lambda_n\circ...\circ\lambda_1$ in $\Gamma$. Their action on $H_{kin}$ is given by
 \begin{align}
 \Pi(m_\ell^2)\psi=\bp_\ell^2\cdot \psi\qquad \Pi(m_\ell s_\ell)\psi=p_\ell^a\cdot \Pi(j_\ell^a)\psi,
 \end{align}
where $p_\ell^a$ are cylindrical functions and $j_\ell^a$ are operators associated with the $IG$-valued holonomy  $H_\ell$ as follows
 \begin{align}
 \label{loopvars}
 &H_\ell =H_{\lambda_n}\cdots H_{\lambda_1}=(u_\ell,-\Ad(u_\ell)\bj_\ell)\\
 &u_\ell=u_{\lambda_n}\cdot u_{\lambda_{n-1}}\cdots u_{\lambda_1}=e^{p_\ell^aJ_a}\qquad 
\bj_\ell=\bj_{\lambda_1}+\Ad(u_{\lambda_1}^\inv)\bj_{\lambda_2}+\ldots+\Ad(u_{\lambda_1}^\inv\cdots u_{\lambda_{n-1}}^\inv)\bj_{\lambda_n}.\nonumber
  \end{align}
A detailed discussion of their action on quantum states and their physical interpretation  is given in the following subsections.

\subsection{The link between combinatorial and loop quantum gravity kinematics}
\label{jqrel}

\subsubsection{Operators in loop quantum gravity and in the combinatorial formalism}

We are now ready to establish the relation between the kinematical operators in the combinatorial formalism and in loop quantum gravity. As discussed in the last subsection, the spaces of quantum states and the kinematical Hilbert spaces  in the two approaches are identical. Moreover,  in both cases functions of the $G$-valued holonomies assigned to the edges of the graph $\Gamma$ act on these spaces by multiplication. However, it remains to clarify the role of the additional structure in the 
combinatorial formalism, the 
 ciliation which establishes a linear ordering of the incident edges at each vertex, and to relate 
  the operators $\bj_\lambda$ and $\bq_\lambda$. While formulas
\eqref{uqdef}, \eqref{jwedef} provide an explicit expression of the associated classical variables in terms of the
triad $e$ and the spin-connection $\omega$, there is a priori no direct link between these variables. However, as we will see in the following, they exhibit a clear and physically intuitive relation
at the quantum level.

\begin{figure}[h]
\psfrag{l}{$\lambda$}
\psfrag{lb}{$\bar\lambda$}
\psfrag{lbs}{$\bar\lambda_s$}
\psfrag{lbt}{$\bar\lambda_t$}
\centering
\includegraphics[scale=0.8]{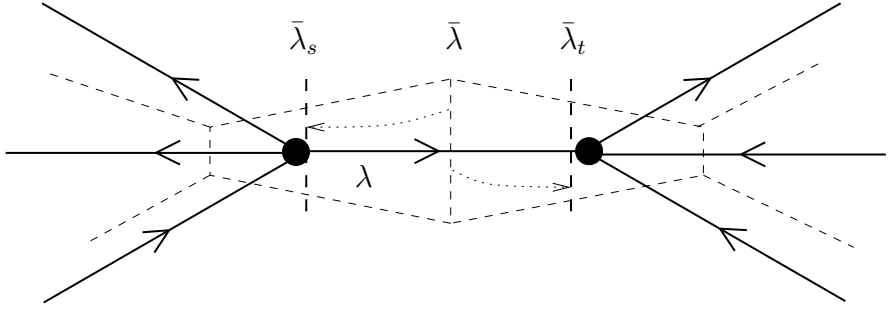}
\caption{\small{Geometrical construction of the operators $\bq_{\lambda,s}$ and $\bq_{\lambda,t}$. We consider
a graph (thick plain lines) and its dual (thin dashed lines): $\lambda$ is the edge between the two vertices.
The operators $\bq_{\lambda,s}$ (resp. $\bq_{\lambda,t}$) are obtained by moving $\bar\lambda$ towards the starting (resp. end) 
point of $\lambda$ and are associated to the dual edges $\lambda_s$ (resp. $\lambda_t$).}}
\label{dualmove}
\end{figure}

We start by determining how  the operators $\bj_\lambda$ in the combinatorial formalism can be understood from the viewpoint of loop quantum gravity. 
For that purpose, we consider the dual $\bar\Gamma$ of the graph $\Gamma$ and the associated operators 
$\bq_{\bar\lambda}$ obtained by integrating the triad over the dual edges $\bar\lambda$ as in \eqref{uqdef}. 
We orient the dual graph in such a way that the intersection number of $\lambda$ and $\bar\lambda$ is $+1$.
As the edges $\lambda$ and $\bar\lambda$ generically cross at a point of $\lambda$, this does not give rise immediately to a well-defined representation of the operators 
$q_{\bar\lambda}$ on the space of cylindrical function $\cif(G^{|E_\Gamma|})$. However, such a representation is obtained if one considers the operators $\bq_{\bar\lambda}$  in the limit where the intersection point of the edge $\lambda$ and its dual edge 
$\bar\lambda$  is moved towards the starting point $s(\lambda)$ or the endpoint $t(\lambda)$, as illustrated  in Fig.~\ref{dualmove}. 

Denoting the associated operators, respectively, by  $\bq_{\lambda,s}$ and $\bq_{\lambda,t}$
and using formula  \eqref{pathbrack}, we then  find that their action on the space of cylindrical functions
is well-defined and given by the left and right-invariant vector fields on $G^{|E_\Gamma|}$
\begin{align}
\label{lambdlim}
\Pi(q_{\lambda,s}^a)\psi=iR_\lambda^a \psi\qquad \Pi(q_{\lambda,t}^a)\psi=-iL_\lambda^a \psi.
\end{align}
Comparing these formulae with expression \eqref{jact} for the action of the operators $\bj_\lambda$, we find that we can identify $\bj_\lambda$ with a certain linear combinations of the operators $\bq_{\lambda,s}$, $\bq_{\lambda,t}$ as follows:
\begin{align}
\label{jqform}
\bj_\lambda=-\bq_{\lambda,s}- \!\!\!\!\!\!\!\!\!\sum_{\tau\in S^+(s(\lambda))}  \!\!\!\!\!\!\!\bq_{\tau,s} + \!\!\!\!\sum_{\tau\in T^+(s(\lambda))}  \!\!\!\!\!\!\!\bq_{\tau,t}+\Ad(u_\lambda^\inv)\left(\sum_{\tau\in S^+(t(\lambda))}  \!\!\!\!\!\!\!\bq_{\tau,s} - \!\!\!\!\sum_{\tau\in T^+(t(\lambda))}  \!\!\!\!\!\!\!\bq_{\tau,t}\right).
\end{align}
This identification will  provide us with a clear geometrical interpretation of the operators $\bj_\lambda$ and their relation 
to the loop quantum gravity variables $\bq_\lambda$. Moreover, it sheds light on the role of the cilia in the two quantisation formalisms.  To see this, we
 consider the following path $\gamma_\lambda$ in the union 
$\Gamma\cup\bar\Gamma$ of the graph $\Gamma$ and depicted  in Fig.~\ref{jdual}: 
\begin{figure}[h]
\psfrag{qs}{$-q_s$}
\psfrag{qt}{$q_t$}
\psfrag{eq}{$q_{\gamma_\lambda}=-q_s+u_\lambda^{-1}q_tu_\lambda$}
\psfrag{u-1}{$u_\lambda^{-1}$}
\psfrag{u}{$u_\lambda$}
\centering
\includegraphics[scale=0.6]{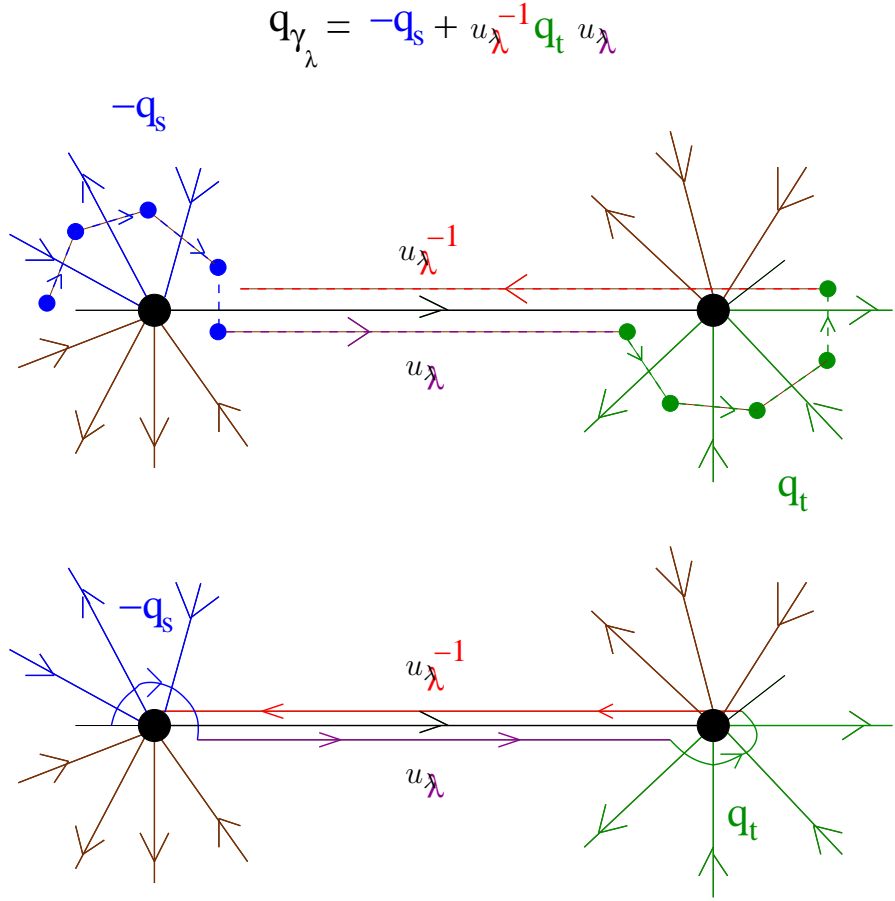}
\caption{\small{The construction of the path $\gamma_\lambda$: the edge $\lambda$ is represented by a black line. The four components of the path $\gamma_\lambda$ and the associated dual edges are depicted in blue, green, purple and red. Other edges incident at the starting and target vertex of $\lambda$ are brown. Cilia
are represented by thin black lines at the vertices and the orientation of the surface is anti-clockwise.}}
\label{jdual}
\end{figure}
\begin{enumerate} 
\renewcommand{\theenumi}{\roman{enumi}}
\renewcommand{\labelenumi}{(\theenumi)}
\item $\gamma_\lambda$
starts at the cilium at the vertex $s(\lambda)$ and goes along the edges of the dual graph $\bar\Gamma$ against the orientation at $s(\lambda)$ until the path crosses the edge $\lambda$ (the blue path in Fig.~\ref{jdual});
\item it continues along $\lambda$ to the vertex $t(\lambda)$ (the purple path in Fig.~\ref{jdual});
\item it goes along the edges of the dual graph in the sense of the orientation at $t(\lambda)$ until 
the path arrives at the cilium at $t(\lambda)$ (the green path in Fig.~\ref{jdual});
\item it goes back along the edge $\lambda$ to the cilium at the starting point $s(\lambda)$ and closes there  (the red path in Fig.~\ref{jdual}).
\end{enumerate}
Note that the resulting  loop goes around the two vertices of $\lambda$ with the associated cilia, and that these cilia together with the orientation
of $S$ determine which of the edges of the dual graph are contained in the loop $\gamma_\lambda$.

Let us now compute the $IG$ valued holonomy $H_{\gamma_\lambda}$ of the path $\gamma_\lambda$. Using 
the group multiplication law \eqref{gmult} and taking into account the orientation of the dual edges,
we find that this holonomy is given by 
\begin{align}\label{Hlambda}
H_{\gamma_\lambda}=(1,\bq_{\gamma_\lambda})=&(u_\lambda^\inv,0)\cdot(1,  \!\!\!\!\!\!\!\sum_{\tau\in S^+(t(\lambda))} \!\!\!\!\!\!\! \bq_{\bar \tau} + \!\!\!\!\!\sum_{\tau\in T^+(t(\lambda))}  \!\!\!\!\!\!\!\bq_{-\bar\tau})
\cdot (u_\lambda,0)\cdot(1, - \!\!\!\!\!\!\!\sum_{\tau\in S^+(s(\lambda))} \!\!\!\!\!\!\!\bq_{\bar \tau} - \!\!\!\!\!\sum_{\tau\in T^+(s(\lambda))}  \!\!\!\!\!\!\!\bq_{-\bar \tau})\;.
\end{align}
The result is given as the product of four terms associated to the different components of the path $\gamma_\lambda$:
\begin{enumerate} 
\renewcommand{\theenumi}{\roman{enumi}}
\renewcommand{\labelenumi}{(\theenumi)}
\item the first one (on the right in  \eqref{Hlambda})
corresponds to the sum over all vectors $\bq_{\bar\tau}$ associated to the duals of edges $\tau$ incident at the starting point of $\lambda$ and of higher order than $\lambda$, taking into account their orientations (the blue arc in Fig.~\ref{jdual});
\item the second term $(u_\lambda,0)$ corresponds to the $G$-holonomy along $\lambda$ (the purple  line in Fig.~\ref{jdual});
\item the third term corresponds to the sum over the vectors $\bq_{\bar \tau}$ for the duals of  edges $\tau$ incident at the target vertex of $\lambda$ and of higher 
order than $\lambda$ (the green arc in Fig.~\ref{jdual});
\item the last term,  $(u_\lambda^\inv,0)$, corresponds to the $G$-holonomy 
along $-\lambda$ (the red  line in Fig.~\ref{jdual}).
\end{enumerate}
We now consider the operator associated to the translational part  $\bq_{\gamma_\lambda}$ of this holonomy.
After moving $\bar\lambda$ and the duals of all other edges incident at the starting point $s(\lambda)$ towards $s(\lambda)$ and the duals of all other edges incident at the $t(\lambda)$ towards $t(\lambda)$ as shown in Fig. \ref{jdual}, formula \eqref{lambdlim} implies that the action of this operator on the states is given by
\begin{align}
\label{jform}
\Pi(q^a_{\gamma_\lambda})\psi=-iR_\lambda^a \psi-\!\!\!\!\!\!\!\!\!\sum_{\tau\in S^+(s(\lambda))} \!\!\!\!\!\!\!iR_\tau^a\psi -\!\!\!\!\!\!\!\!\!\sum_{\tau\in T^+(s(\lambda))} \!\!\!\!\!\!\!iL_\tau^a\psi +i\Ad(u_\lambda^\inv)^{a}_{\;\;b}\left(\sum_{\tau\in S^+(t(\lambda))} \!\!\!\!\!\!\!R_\tau^b\psi +\!\!\!\!\!\!\!\!\!\sum_{\tau\in T^+(t(\lambda))} \!\!\!\!\!\!\!L_\tau^b\psi \right),
\end{align}
which agrees with equation  \eqref{jact} for the action of $\bj_\lambda$. Hence, we can identify the operators $\bj_\lambda$ in the combinatorial formalism with the loop quantum gravity  operator $\bq_{\gamma_\lambda}$  for the path $\gamma_\lambda$ in the limit where the edges of the dual graph are moved towards the starting and target vertex of $\Gamma$.

\subsubsection{Physical interpretation}

Equation \eqref{jform} is one of the core results of our paper and  provides a clear geometrical interpretation of the kinematical operator $\bj_\lambda$ and its relation to the loop quantum gravity variables $\bq_\lambda$.
The definition \eqref{uqdef} of the classical variables  associated with the operators $\bq_\gamma$  suggests an interpretation of the operators $\bq_\gamma$ as a relative position vector of the ends of the path $\gamma$. 
With this interpretation the terms
\begin{align}
\bq_s=\sum_{\tau\in S^+(s(\lambda))} \!\!\!\!\!\!\!\bq_{\bar \tau} + \!\!\!\!\!\sum_{\tau\in T^+(s(\lambda))}  \!\!\!\!\!\!\!\bq_{-\bar \tau}\qquad \bq_t=\sum_{\tau\in S^+(t(\lambda))} \!\!\!\!\!\!\! \bq_{\bar \tau} + \!\!\!\!\!\sum_{\tau\in T^+(t(\lambda))}  \!\!\!\!\!\!\!\bq_{-\bar\tau}
\end{align}
in \eqref{Hlambda} which are depicted in Fig.~\ref{jdual} can be viewed as the relative position vectors of  the intersection point $\lambda\cap\bar\lambda$ with respect to the cilia at the starting and target vertex of $\lambda$. 

In the limit where the dual edges are moved towards the starting and target vertex of $\lambda$  they can be interpreted as, respectively, the relative position vectors of $s(\lambda)$ and $t(\lambda)$ with respect to the cilia at these vertices. 
The $G$-valued holonomy $u_\lambda$ has the interpretation of a Lorentz transformation or rotation relating the two reference frames associated with the starting and target vertex of $\lambda$. Conjugating the relative position vector $\bq_t$ at $t(\lambda)$  with the inverse of this holonomy therefore corresponds to transporting it into the reference frame associated with the starting vertex $s(\lambda)$. The operator $\bq_{\gamma_\lambda}$ is then obtained by subtracting $\bq_s$ from $\Ad(u_\lambda^\inv)\bq_t$. It can therefore be viewed as a relative position vector of the edge ends $s(\lambda)$ and $t(\lambda)$ in the reference frame associated with  $s(\lambda)$.

The relation between the operators $\bq_\lambda$ and $\bj_\lambda$ also sheds light on the role of the cilium in the combinatorial and the loop formalisms: The addition of cilia at each vertex corresponds to the choice of a reference point which allows one to consistently assign a position vector to each edge incident at the vertex. It therefore enters the definition of the variables $\bj_\lambda$ which give the relative position of the starting point and the endpoint of $\lambda$ in the reference frame associated with its starting point. Note that this interpretation is also supported by the transformation of the variables  $\bj_\lambda$, $\bq_\lambda$ under the reversal of edges given in \eqref{frinvert}: While the position vectors $\bq_\lambda$ acquire a minus sign, the operators $\bj_\lambda$ acquire a minus sign and  are  multiplied with a factor $\Ad(u_\lambda)$, which describes their transport in the reference frame associated with the target vertex.

\subsubsection{The case of a loop}

\begin{figure}[h]
\psfrag{l}{$\lambda$}
\centering
\includegraphics[scale=0.5]{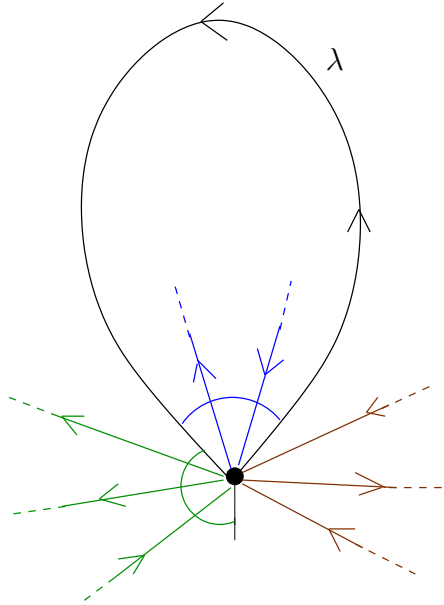}
\caption{\small{Example of a loop $\lambda$ attached to a vertex. The part which does not contain the cilium defines
the inside of $\lambda$. The blue edges inside (resp.~the green edges outside) the loop contribute to the spin-vector (resp.~orbital momentum) associated to the loop $\lambda$. The edges depicted in brown do not contribute to the variable $\bj_\lambda$.}}
\label{dualangle}
\end{figure}

To deepen the understanding of the relation between loop quantum gravity operators and combinatorial operators and their physical interpretation,  it is instructive to consider the situation where $\lambda$ is a loop as depicted in Fig.~\ref{dualangle}. For notational convenience  we assume its ends to be ordered such that  $O(\lambda,s)<O(\lambda,t)$. 
The expression \eqref{jqform} for the associated operator $\bj_\lambda$ then simplifies and can be written as a sum $\bj_\lambda=\bs_\lambda+\bel_\lambda$ with
\begin{eqnarray}
\bs_\lambda  & = & -\bq_{\lambda,s}-  \!\!\!\!\!\!\!\! \sum_{\tau\in S^+(s(\lambda))\cap S^-(t(\lambda))} \!\!\!\!\! \!\!\!\!\!\bq_{\tau,s} + \!\!\!\sum_{\tau\in T^+(s(\lambda))\cap T^-(t(\lambda))}  \!\!\!\!\! \!\!\!\!\bq_{\tau,t} \label{spinlopdef}\\
\bel_\lambda & = & -(1-\Ad(u_\lambda^\inv))\left(\sum_{\tau\in S^+(t(\lambda))}  \!\!\!\!\!\!\!\bq_{\tau,s} -  \!\!\!\!\!\!\sum_{\tau\in T^+(t(\lambda))}  \!\!\!\!\!\!\!\bq_{\tau,t}\right).\label{angloopdef}
\end{eqnarray}
By considering these two terms illustrated in Fig. \ref{dualangle} we find  that the vector $\bs_\lambda$ corresponds to the contribution of the edges ``inside" the loop $\lambda$ (the blue edges in Fig.~\ref{dualangle}
and  $\bel_\lambda$ to the one of the edges ``outside" the loop  $\lambda$ and of higher order than $t(\lambda)$ (the green edges in Fig.~\ref{dualangle}). The notions of  ``inside" and ``outside" are provided by the cilium:  the ``outside"  of the loop is the component of the surface $S$ which contains the cilium when $S$ is cut along the loop.
We interpret these two contributions by relating them to the kinematics of 
 particles in  three-dimensional gravity, which  have been discussed extensively by many authors \cite{LPR,we1,Hooft, DJ,djt, SG,  Matschull1, LM1,LM2, Matschpart,NP,K1}.

At the kinematical level, a relativistic particle moving in 
 three-dimensional Euclidean or  Minkowski space is characterised by a position three-vector $\bx$ and its momentum three-vector $\bp$. The kinematical observables are the momentum three-vector $\bp$ together with the  total angular momentum three-vector $\bj$. They form a Poisson algebra which reproduces the three-dimensional Euclidean or Poincar\'e algebra.  The mass $m$
and the spin $s$ of the particle are given  by the Casimir functions of the Poisson algebra $\bp^2=m^2$ and $\bp \cdot \bj = ms$.  As a consequence, the  total angular
momentum three-vector $\bj$ decomposes naturally  into its longitudinal component $\bs$ with respect to $\bp$ and its orbital angular momentum $\bel$
\begin{align}
\label{classjrel}
\bj=\bs+\bel=\tfrac{s}{m}\bp+\bx\wedge\bp.
\end{align}
It has been shown that in the presence of gravitational interaction the momenta of particle become group valued. Equation \eqref{classjrel} is modified and approaches the non-gravitational form \eqref{classjrel} in the low-mass limit $\bp^2\rightarrow 0$
\begin{align}
\label{gravloopdef}
\bj=\tfrac{s}{m}\bp+(1-\Ad(e^{-p^aJ_a}))\bx= \tfrac{s}{m}\bp+\bp\wedge \bx+O(\bp^2).
\end{align} 
To exhibit the link with the kinematics of a classical particle, we parametrise the $G$-holonomy of the loop $\lambda\in E_\Gamma$ in terms of a three-vector $\bp_\lambda$ 
\begin{align}
u_\lambda=e^{p^a_\lambda J_a} ,\qquad  \bp_\lambda^2=m_\lambda^2.
\end{align}
A short calculation involving \eqref{spinlopdef}, \eqref{angloopdef} then yields
\begin{align}
\bp_\lambda\cdot \bj_\lambda=\bp_\lambda\cdot \bs_\lambda = m_\lambda s_\lambda \qquad \bp_\lambda\bel_\lambda=0
\end{align}
This implies that only the inside edges between the two ends of the loop $\lambda$ (the blue edges in Fig.~\ref{dualangle}) contribute to the spin of a loop $\lambda$. The projection of
the sum over the position vectors of these internal edges in the direction of $\bp_\lambda$  can be viewed as an internal angle, which  generalises the deficit
angle arising in spacetimes with particles.  The associated quantity $s_\lambda$ therefore defines an internal 
angular momentum or spin. The component $\bel_\lambda$ defined in \eqref{angloopdef} which arises from the external edges (green in Fig.~\ref{dualangle}) is necessarily orthogonal to the momentum $\bp_\lambda$
and therefore contributes only to the orbital angular momentum in  \eqref{gravloopdef}. 
The sum over the position vectors of the external edges can therefore be viewed as an external position vector for the loop with respect to the cilium at its vertex.

\section{Quantum double symmetries}

\label{qudoublesymms}

\subsection{The quantum double $D(G)$ in three-dimensional gravity}
\label{qudoublesect}

In this section we derive the second  core result of our paper:
We demonstrate how quantum group symmetries arise in  three-dimensional loop quantum gravity and the combinatorial quantisation formalism. The relevant quantum groups are the quantum doubles $D(G)$ of the three-dimensional Lorentz and rotation group.
The role of the quantum groups in the combinatorial formalism is well-understood for the case where 
the graph  $\Gamma$ is a minimal simplicial decomposition of the
surface $S$, i.~e.~a set of generators of the fundamental group $\pi_1(S)$ \cite{AGSI,AGSII,BNR,we2,BMS}. 

However, the situation is less clear when $\Gamma$ is a general graph on the surface $S$, which is the case generically in the loop formalism and  in spinfoam models. 
In three-dimensional gravity with vanishing cosmological constant, no direct evidence of quantum group symmetry has been detected in the loop and spin foam approaches.  
The evaluation of link invariants for the quantum double $SU(2)$ and  their relation to the Ponzano-Regge model are investigated in \cite{LPR, BPR}. However, the 
 role of quantum groups remains indirect and implicit in these papers. In particular,  they do not shed light on the general relation between quantum groups and the generic building blocks of these formalisms, graphs and spin network functions. 
 
 This raises the question if quantum group symmetries are generic features of three-dimensional quantum gravity or rather mathematical tools within the combinatorial approach based on a minimal simplicial decomposition.  In this section, we demonstrate that quantum group symmetries appear as a generic feature of three-dimensional quantum gravity with vanishing cosmological constant and are also present in the loop formalism. We show that the quantum double $D(G)$ acts naturally on the space of cylindrical functions for general graphs $\Gamma$.  More concretely, we demonstrate that each closed, non-selfintersecting loop in $\Gamma$ gives rise to a representation of the quantum double on the space of cylindrical functions and derive explicit expressions for these representations in Sect.~\ref{explquact}. Moreover, we show in Sect.~\ref{kinhilbquant} that there is a remnant of these representations on the kinematical Hilbert space
  which is directly related to the fundamental kinematical observables studied in the previous section.

We start with a definition of the quantum double $D(G)$, also called the Drinfeld double. For a brief summary of its representation theory we refer the reader to appendix \ref{qurep}.
The quantum double $D(G)$ is a quasi-triangular ribbon Hopf algebra which can be identified (as a vector space) with 
the tensor product
\be
D(G) \; \equiv \; D(F(G)) = F(G) \otimes \mathbb C(G) 
\ee
of the space  $F(G)$ of functions on $G$ and the group algebra $\mathbb C(G)$. 
Here we follow the presentation in \cite{we2}
and work with an alternative formulation which is advantageous as it exhibits explicitly the close link between the classical and quantised theory. 
In this description, first given in \cite{kM}, the
 quantum double $D(G)$ is  formulated in terms of continuous
functions on $G\times G$. To exhibit its structure as a quasi-triangular ribbon-Hopf algebra  it is necessary to include certain Dirac delta-distributions $f\otimes\delta_g$, which are not elements of the space of
continuous functions $C_0(G\times G)$ but can be included by adjoining them. The Hopf algebra structure of $D(G)$ is then given as follows:
\ba
\label{prod}
&& \text{Product} \; : \;
(F_1\bullet F_2)(u,v):=\int_G F_1(u,z)\,F_2(z^{-1}uz, z^\inv v)\,dz,
 \\
&& \text{Coproduct} \; : \; (\Delta F)(u_1, v_1;u_2,v_2)=F(u_1u_2,v_1)\,\delta_{v_1}(v_2), \\
&& \text{Antipode} \; : \; (S F)(u,v)=F(v^{-1}u^{-1}v, v^\inv), \\
&& \text{Unit} \; : \; 1(u,v)=\delta_e(v),\\
&& \text{Counit} \; : \; \varepsilon(F)=\int_G F(e,v)\,dv, \\
&& \text{Star structure} \; : \;F^*(u,v)=\overline{F(v^{-1}uv, v^\inv)}\, .\label{starstruct}
\ea
For the singular elements $f\otimes\delta_g$, expressions \eqref{prod} to \eqref{starstruct} take the form
\ba
\label{prod2}
&& \text{Product} \; : \;
( f_1\otimes\delta_{g_1})\bullet(f_2\otimes\delta_{g_2})=(f_1\cdot f_2\circ\Ad_{g_1^\inv})\otimes \delta_{g_1g_2}, \\
&& \text{Coproduct} \; : \; \Delta( f\otimes\delta_g)(u_1,v_1;u_2,v_2)=f(u_1u_2)\,\delta_g({v_1})\delta_g(v_2), \label{coprod}\\
&& \text{Antipode} \; : \; S( f\otimes\delta_g)(u,v)=f(v^{-1}u^{-1}v)\delta_{g^\inv}(v),\\
&& \text{Counit} \; : \; \varepsilon(f\otimes\delta_g)=f(e),\\
&& \text{Star structure} \; : \;(f\otimes\delta_g)^*=
(\overline{f}\circ\Ad_{g^\inv})\otimes\delta_{g{-1}}.\label{starstruct2}
\ea
The Hopf algebra  $D(G)$ is quasi-triangular with $R$-matrices, $R^{(\pm)} \in D(G)^{\otimes 2}$,  which are the quantum counterparts, respectively, of the classical $r$-matrix \eqref{rmat} and minus its flip
\begin{align}
&R^{(+)}(u_1,v_1;u_2,v_2) = \delta_e(v_1)\delta_e(u_1v_2^{-1})\qquad R^{(-)}(u_1,v_1;u_2,v_2) = \delta_e(v_2)\delta_e(u_2v_1).
\end{align}
Its ribbon element which satisfies the ribbon relation $\Delta c= (R_{21}\bullet R) \bullet \bigl(  c\otimes c \bigr)$
with the opposite $R$-matrix $R_{21}(u_1,v_1;u_2,v_2):=R(u_2,v_2;u_1,v_1)$ is given by
\begin{align}
\label{randc}
c(u,v) =  \delta_v(u)\,.
\end{align}
It can be shown that $D(G)$ is a deformation (in the sense of Drinfeld) of the classical group algebra
$\mathbb C(IG)$  with the Planck length $\ell_P$ as a deformation parameter. In fact, as an algebra 
$D(G)$ is included into $\mathbb C(IG)$ and the deformation concerns only the co-algebra structures.

\subsection{Quantum double action on the  space of cylindrical functions}

The first indication that the quantum double arises as a symmetry of quantum gravity is the Poisson bracket  \eqref{pathbrack} of the loop variables. Using the formula \eqref{coprod} for the coproduct, we can rewrite this Poisson bracket as 
\begin{align}
\{q^a_\lambda, f\}(u_{\tau_2}u_{\tau_1})=&\frac{d}{dt}|_{t=0} f(u_{\tau_1}e^{tJ_a} u_{\tau_2})\\
=&(\text{id}\otimes -L^a)\circ\Delta f(u_{\tau_1}\otimes u_{\tau_2})=(R^a\otimes \text{id})\circ\Delta f(u_{\tau_1}\otimes u_{\tau_2}).\nonumber
\end{align}
Hence, the coproduct of the quantum double is present already in the Poisson structure of the classical theory and, consequently, also in the action of the associated operators on the kinematical Hilbert space.

However, the role of quantum double symmetries is not limited to this rather indirect manifestation. 
We will now demonstrate that the quantum double $D(G)$ arises naturally as a quantum symmetry also in the loop formulation of the theory and acts on the Hilbert space of the theory. 
More specifically, we will show that each closed, non-selfintersecting loop in the graph $\Gamma$ gives rise to a representation of the quantum double on the space of cylindrical functions $\cif(G^{|E_\Gamma|})$.  As this is one of the core results of our paper and technically rather involved, we will proceed in two steps: We start by illustrating the general structure of these representations.  In Subsection \ref{explquact} we then derive explicit expressions for these representations and discuss their physical interpretation. In Subsection \ref{kinhilbquant} we show how a remnant of this quantum group symmetry manifests itself on the kinematical Hilbert space.

To exhibit the  general structure of these representations, we consider a closed loop $\ell=\lambda_n\circ\lambda_{n-1}\circ\ldots\circ \lambda_1$ in $\Gamma$ which is composed of one or several links $\lambda_1,\ldots,\lambda_n\in E_\Gamma$ and based at a vertex $v=s(\lambda_1)=t(\lambda_n) \in V_\Gamma$.  We assume  $O(\lambda_1,s)< O(\lambda_n,t)$.  For notational convenience we also impose that all edges arising in the loop are oriented in the sense of the loop 
as pictured in Fig. \ref{jloop}.  Moreover, we require that the  loop $\ell$ does not have any self-intersections, i.~e.~that we have
\begin{align}
\lambda_j\cap\lambda_k=\emptyset \;\text{for}\; |k-j|\geq 2, \{k,j\}\neq\{1,n\}\quad \lambda_j\cap\lambda_{j-1}=s(\lambda_j)=t(\lambda_{j-1})\;j=2,...,n
\end{align}
and that none of the edges $\lambda_i$ is a loop unless $n=1$.

We use the notation $H_\ell=(u_\ell,-\Ad(u_\ell)\bj_\ell)$ with $u_\ell$, $\bj_\ell$ as in \eqref{loopvars}. While the action of $u_\ell$ on the space of cylindrical functions is multiplicative, the operator $\bj_\ell$ is derivative. Moreover, as we will show in the next subsection, it generates a group action
$\rho_\ell: G\times G^{|E_\Gamma|}\rightarrow G^{|E_\Gamma|}$:
\begin{align}
\label{jbhkin}
\Pi_\ell(j^a_\ell)\psi=i\{j^a_\ell, \psi\}=i\frac{d}{dt} |_{t=0}\psi\circ\rho_\ell(e^{tJ_a}) \qquad \forall \psi\in\cif(G^{|E_\Gamma|})\,.
\end{align}
It is shown in \cite{we3}, see in particular Lemma 4.2, that a group action $\rho$ of $G$ on a manifold $M$ together with a map $\phi: M\rightarrow G$ satisfying the covariance condition $\Phi(\rho(g)m)=g\cdot \Phi(m)\cdot g^\inv$ $\forall m\in M, g\in G$ gives rise to a representation of $D(G)$ on $\cif(M)$ defined by
\begin{align}
\label{genrep}
\Pi(F)\psi(m)=\int_G d\mu(z)\, F(\Phi(m),z)\cdot \psi\circ\rho(z^\inv)\qquad \forall \psi\in\cif(M).
\end{align}
 In the case at hand, this group action is $\rho_\ell$, the manifold $M=G^{|E_\Gamma|}$ is given by the $G$-holonomies assigned to the edges of the graph $\Gamma$,  and the map $\Phi: G^{|E_\Gamma|}\rightarrow G$ expresses the loop holonomy $u_\ell$ as a product of the edge holonomies $u_{\lambda_i}$
 \begin{align}
 \label{loopphi}
 \Phi:\;(u_1,\ldots, u_{|E_\Gamma|})\mapsto u_\ell=u_{\lambda_n}\cdots u_{\lambda_1}.
 \end{align}
 The covariance condition then takes the form
 \begin{align}
\label{holcond}
\rho_\ell(g) u_\ell=g\cdot u_\ell \cdot g^\inv.
\end{align}
Hence, to demonstrate that the loop $\ell$ in $\Gamma$ gives rise to a representation of the quantum double on the space of cylindrical functions, we need to construct a group action $\rho_\ell: G\times G^{|E_\Gamma|}\rightarrow G^{|E_\Gamma|}$ 
   that satisfies \eqref{jbhkin} and acts on the 
 holonomy $u_\ell$ by 
conjugation. 
It then follows directly from expression \eqref{physscalprod}  for the scalar product and expression \eqref{starstruct2} for the star structure  that this representation is unitary. 
Moreover, formula \eqref{genrep} implies that  the action of the elements $f\otimes\delta_g$ in \eqref{doublerep} takes the particularly simple form
\begin{align}
\label{doublerep2}
\Pi_\ell(f\otimes\delta_g)\psi=f(u_\ell)\cdot \psi\circ\rho_\ell(g^\inv).
\end{align}
In particular, we see that elements $f\otimes 1$ represent  the multiplicative  action of functions of the holonomy $u_\ell$ on 
the space of quantum states, while the elements $1\otimes g$, $g\in G$, exponentiate the action of the operators $\bj_\ell$.

\subsection{ Explicit expressions for the action of the quantum double}

\label{explquact}
We will now construct the group action $\rho_\ell$ for a general non-selfintersecting loop $\ell=\lambda_n\cdots\lambda_1$. 
Due to the close link between  the classical and quantum theories, it is clear that this amounts to exponentiating  the Poisson brackets of  $\bj_\lambda$ with functions $f\in\cif(G^{|E_\Gamma|})$, expressed in terms of the left- and right-invariant vector fields \eqref{compvecfields} and hence will be defined via a graphical procedure similar to the one introduced after \eqref{jbrack}.
However, this  requires replacing sums of vector fields with products of elements of $G^{|E_\Gamma|}$ and one has to demonstrate that there exists an appropriate ordering which gives rise to a group action with the required properties.

To do this, we define explicitly a map $\rho_\ell: G\times G^{|E_\Gamma|}\rightarrow G^{|E_\Gamma|}$ that satisfies \eqref{jbhkin} and then demonstrate that it is a group action, i.~e.~satisfies $\rho_\ell(gh)=\rho_\ell(g)\cdot\rho_\ell(h)$, and that it acts on the holonomy $u_\ell$ by conjugation.
For clarity, we consider separately the following cases: \vspace{-.2cm}
\begin{enumerate} 
\renewcommand{\theenumi}{\roman{enumi}}
\renewcommand{\labelenumi}{(\theenumi)}
\item  the action on edges which have no vertex
in common with the loop $\ell$\vspace{-.2cm}
\item  the action on the edges  $\lambda_1,\ldots,\lambda_n$ which form the loop\vspace{-.2cm}

\item  the action on edges
which have at least one  vertex in common with the loop but do not belong to the loop themselves.\vspace{-.2cm}

\end{enumerate}

{\bf Case (i)}:
It follows directly from \eqref{jact}, \eqref{loopvars} that the operator $\bj_\ell$ acts trivially on the variables $u_\tau$ of all edges $\tau$ that do not have a vertex in common with $\ell$. This suggests that these group elements should transform trivially under $\rho_\ell$.

{\bf Case (ii)}: To determine the action of $\rho_\ell$ on the edges $\lambda_1,\ldots,\lambda_n$ in the
loop, we start by considering  the extreme edges $\lambda_1$ and $\lambda_n$. 
Using expression \eqref{jact} together with \eqref{loopvars}, we find
\begin{align}
&-i\Pi(j^a_\ell)f_{\lambda_1}=\begin{cases} 0 & \;\;\;\;\;\;\;\;\;\;\;\;\;\;\;\;\;\;\;\;\;\;\;\;\,\text{if}\;
O(\lambda_1,t)>O(\lambda_2,s)\\
-R^a f_{\lambda_1} &  \;\;\;\;\;\;\;\;\;\;\;\;\;\;\;\;\;\;\;\;\;\;\;\;\,
\text{if}\;O(\lambda_1,t)<O(\lambda_2,s) \end{cases}\\
& -i\Pi(j^a_\ell)f_{\lambda_n}=\begin{cases} -L^af_{\lambda_n} & \text{if}\;O(\lambda_n,s)>O(\lambda_{n-1},t)\\
-(1-\Ad(u_l^\inv))^{a}_{\;\;b}L^bf_{\lambda_n} & \text{if}\;O(\lambda_n,s)<O(\lambda_{n-1},t)\end{cases}\nonumber
\end{align}
While exponentiating  the first three terms is straightforward, the last involves an ordering ambiguity for the factors. Supposing that identity \eqref{holcond} is satisfied, we see that in order to have the group action property  $\rho_\ell(gh)=\rho_\ell(g)\cdot \rho_\ell(h)$
 the holonomies $u_{\lambda_1}, u_{\lambda_n}$ have to transform  as
\begin{align}
\label{rhol1act}
&u_{\lambda_1}\mapsto \begin{cases} u_{\lambda_1} & \;\text{if}\;O(\lambda_1,t)>O(\lambda_2,s)\\
 u_{\lambda_1}\cdot g^{-1} & \; \text{if}\;O(\lambda_1,t)<O(\lambda_2,s)
\end{cases} \\
&u_{\lambda_n}\mapsto \begin{cases} g\cdot u_{\lambda_n} & \text{if}\;O(\lambda_n,s)>O(\lambda_{n-1},t)\\
[g,u_\ell] \cdot u_{\lambda_n} & \text{if}\;O(\lambda_n,s)<O(\lambda_{n-1},t)\nonumber
\end{cases}
\end{align}
where $[a,b]=a\cdot b\cdot a^\inv \cdot b^\inv$ is the group commutator of $G$. 
An analogous reasoning for the other edges $\lambda_k$,  $k=2,\ldots, n-1$ yields
\begin{align}
\label{rhol2act}
&u_{\lambda_k}\mapsto\\
& \begin{cases} u_{\lambda_k} & \text{if}\;O(\lambda_k,s)>O(\lambda_{k-1},t)\;\text{and} \;O(\lambda_{k+1},s)>O(\lambda_k,t)\\
u_{\lambda_k} & \text{if}\;O(\lambda_k,s)<O(\lambda_{k-1},t)\;\text{and} \;O(\lambda_{k+1},s)<O(\lambda_k,t)\\
u_{\lambda_k}\cdot(u_{\lambda_{k-1}}\!\!\!\!\!\!\cdots u_{\lambda_1})g(u_{\lambda_{k-1}}\!\!\!\!\!\!\cdots u_{\lambda_1})^\inv & \text{if}\;O(\lambda_k,s)>O(\lambda_{k-1},t)\;\text{and} \;O(\lambda_{k+1},s)<O(\lambda_k,t)\\
u_{\lambda_k}\cdot(u_{\lambda_{k-1}}\!\!\!\!\!\!\cdots u_{\lambda_1})g^\inv(u_{\lambda_{k-1}}\!\!\!\!\!\!\cdots u_{\lambda_1})^\inv & \text{if}\;O(\lambda_k,s)<O(\lambda_{k-1},t)\;\text{and} \;O(\lambda_{k+1},s)>O(\lambda_k,t).
\end{cases}\nonumber
\end{align}
It can then be shown by a straightforward calculation that the map $\rho_\ell$ defined by \eqref{rhol1act}, \eqref{rhol2act} acts on ordered products of the edge holonomies $u_{\lambda_i}$ according to \begin{eqnarray}
\label{rhoacttot}
u_\ell &\mapsto& g\cdot u_\ell\cdot g^\inv\\
u_{\lambda_{k}}\cdots u_{\lambda_1} &\mapsto&
\begin{cases} u_{\lambda_{k}}\cdots u_{\lambda_1} & \text{if}\;O(\lambda_{k+1},s)<O(\lambda_k,t)\\
u_{\lambda_{k}}\cdots u_{\lambda_1}\cdot g^\inv & \text{if}\;O(\lambda_{k+1},s)>O(\lambda_k,t)
\end{cases}\qquad k=1,\ldots, n-1.\nonumber
\end{eqnarray}

{\bf Case (iii)}: We distinguish two cases: edges that start or end at the starting vertex of $\ell$ and edges that start or end at other vertices $s(\lambda_{k+1})=t(\lambda_k)$, $k\neq n$. While the relative order of the incident edges in the loop is fixed in the former, it is not in the latter, and we have to consider separately the situation where $O(\lambda_{k+1}, s)>O(\lambda_k,t)$ and $O(\lambda_{k+1}, s)<O(\lambda_k,t)$. 

We start by considering an edge $\tau$ 
starting at the vertex $s(\lambda_{k+1})=t(\lambda_{k})$ with
$k\neq n$,  where the order of the edges $\lambda_{k+1}$, $\lambda_{k}$ is $O(\lambda_{k},t)>O(\lambda_{k+1},s)$. Using again formulas  \eqref{jact},\eqref{loopvars},
we determine the action of $\bj_\ell$ on $u_\tau$ and find that the map $\rho_\ell$ should act on these holonomies according to
\begin{align}
\label{genedgeact}
u_\tau\!\mapsto\!\begin{cases} u_\tau & \text{if}\;O(\tau,s)\!<\!O(\lambda_{k+1},s)\;\text{or}\; O(\tau,s)\!>\!O(\lambda_{k},t)\\
u_\tau\cdot (u_{\lambda_{k}}\!\!\!\cdots u_{\lambda_1}) g^\inv (u_{\lambda_{k}}\!\!\!\cdots u_{\lambda_1})^\inv & \text{if}\;O(\lambda_{k+1},s)\!<\!O(\tau,s)\!<\!O(\lambda_{k}, t).\end{cases}
\end{align}
Analogously, we find for an  edge $\tau$ starting at the vertex $s(\lambda_{k+1})=t(\lambda_{k})$ with
 $k\neq n$, where the order of the incident edges in the loop is $O(\lambda_{k},t)<O(\lambda_{k+1},s)$
\begin{align}
\label{genedgeact2}
&u_\tau\mapsto\begin{cases} u_\tau & \text{if}\;O(\tau,s)<O(\lambda_{k},t)\;\text{or}\; O(\tau,s)>O(\lambda_{k+1},s)\\
u_\tau\cdot (u_{\lambda_{k}}\!\cdots u_{\lambda_1}) g (u_{\lambda_{k}}\!\!\!\cdots u_{\lambda_1})^\inv & \text{if}\;O(\lambda_{k},t)<O(\tau,s)<O(\lambda_{k+1}, s).\end{cases}
\end{align}

The corresponding expressions for  an edge $\tau$ starting at the vertex $s(\lambda_1)=t(\lambda_{n})$ 
are analogous but involve an additional contribution for the edges of higher order than $\lambda_n$.
: \begin{align}
\label{genedgeact3}
u_\tau\mapsto\begin{cases} u_\tau & O(\tau,s)<O(\lambda_{1},s)\\
u_\tau\cdot g^\inv & O(\lambda_{1},s)<O(\tau,s)<O(\lambda_{n}, t)\\
u_\tau\cdot[u_\ell,g] & O(\tau,s)>O(\lambda_n, t)\end{cases}.
\end{align}
The action of $\rho_\ell$ on
the holonomies associated to  edges  that end at the vertices in the loop 
is obtained by exchanging right-multiplication with group elements $a\in G$ with left-multiplication by 
$a^\inv$ in expressions \eqref{genedgeact} to \eqref{genedgeact3}. The corresponding expressions for
loops based on these vertices are then obtained 
by applying this prescription to both ends of the loop.

This concludes our discussion of the different cases. Equations \eqref{rhol1act} to \eqref{genedgeact3} provide an explicit definition of $\rho_\ell$ through its action on the holonomies of all edges in $\Gamma$. Formula \eqref{rhoacttot} demonstrates that it satisfies the covariance condition. By differentiating \eqref{rhol1act} to \eqref{genedgeact3} and comparing the result with the action of the loop operator $\bj_\ell$ given by  \eqref{loopvars}, \eqref{jact} we verify \eqref{jbhkin} and find that the action of $\bj_\ell$ on the cylindrical functions is indeed the infinitesimal version of the map $\rho_\ell$.  It remains to 
show that $\rho_\ell$ is a group action. This can be shown by a straightforward but somewhat lengthy calculation using expressions \eqref{rhol1act} to \eqref{genedgeact3}.
Hence,  we have demonstrated that the action of the loop operator $\bj_\ell$ gives rise to a group action $\rho_\ell: G\times G^{|E_\Gamma|}\rightarrow G^{|E_\Gamma|}$ with the required invariance properties and defines a representation of the quantum double $D(G)$.

\begin{figure}[h]
\centering
\psfrag{l1}{$\lambda_1$}
\psfrag{l2}{$\lambda_2$}
\psfrag{l3}{$\lambda_3$}
\psfrag{ln}{$\lambda_n$}
\psfrag{lnm}{$\lambda_{n-1}$}
\psfrag{v}{$v$}
     \includegraphics[scale=0.6]{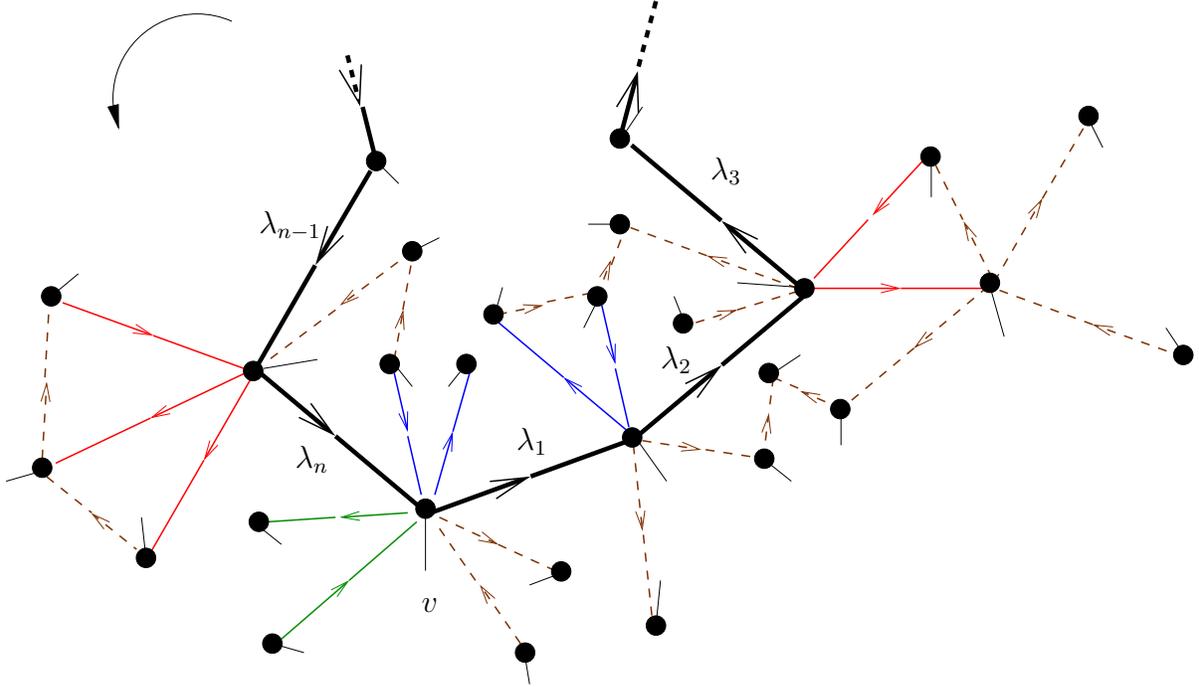}
    \caption{\small{ Illustration of the group action $\rho_\ell$ associated to the closed, non-self-intersecting loop $\ell=\lambda_n\cdots\lambda_1$. The oriented loop $\ell$ is represented by a solid black line and the cilia at its vertices by thin black lines.  Edges whose variables transform trivially under $\rho_\ell$ are depicted as brown dashed lines. Red and blue edges correspond to the non-trivial transformations in, respectively, \eqref{genedgeact} and \eqref{genedgeact2}. The variables associated with the green edges at the starting vertex of $\ell$ transform according to the last line in \eqref{genedgeact3}.}}
  \label{jloop}
\end{figure}

This demonstrates that  each closed, non-selfintersecting loop $\ell$ in the graph $\Gamma$ gives rise to a representation of the quantum double $D(G)$ on the space of cylindrical functions for $\Gamma$ defined by \eqref{genrep}, \eqref{doublerep2}.
Moreover, these representations have a clear 
 geometrical interpretation which encodes the topology and the orientation of the graph $\Gamma$: Holonomies $u_\lambda$  transform trivially if the associated edges $\lambda$ do not intersect the loop.  
 The holonomies associated with the edges $\lambda_1,\ldots,\lambda_n$ in the loop transform non-trivially if and only if the relative order of consecutive edges at the starting and endpoint changes, i.~e.~if the associated cilia point in different directions with respect to the orientation of the loop.  
 Expressions \eqref{genedgeact}, \eqref{genedgeact2} imply that
  holonomies of edges $\tau$ which are not part of the loop but have a vertex $s(\lambda_k)=t(\lambda_{k-1})$, $k\neq 1$ in common with it, transform nontrivially if and only if they lie between the two edges of the loop touching this vertex with respect to the ordering. 
 Defining the ``inside" and ``outside" of a loop with respect to the cilium at each vertex as in the paragraph following \eqref{angloopdef}, we find again that only the inner edges at each vertex are affected by the loop operator $\bj_\ell$ and the associated group action $\rho_\ell$.
 At the starting vertex $s(\lambda_1)=t(\lambda_n)$ there is an additional contribution for edges of higher order than $\lambda_n$. These cases are illustrated in the Fig. \ref{jloop}.

By differentiating expressions \eqref{rhol1act} to \eqref{genedgeact3}, one obtains a pattern similar to the one for a single-edge  loop  in Sect.~\ref{jqrel}.  Expressing the operator $\bj_\ell$ in terms of the operators $\bq_\lambda$ associated to the edges of the dual graph and moving these dual edges towards the starting and target vertices of the edges $\lambda_i$,  we find that $\bj_\ell$ is given as a sum $\bj_\ell=\bs_\ell+\bel_\ell$ with
\begin{align}
\label{angmomgenloop}
&\bel_\ell=-(1-\Ad(u_\ell^\inv))\left(\sum_{\tau\in S^+(t(\lambda_n))} \!\!\!\!\!\bq_{\tau,s}- \!\!\!\!\!\sum_{\tau\in T^+(t(\lambda_n))}  \!\!\!\!\!\bq_{\tau,t}\right)\\
&\bs_\ell= \sum_{i=0}^{n-1} \varepsilon_i\,\Ad(u_{\lambda_{i}}^\inv\cdots u_{\lambda_1}^\inv)\left( \sum_{\tau\in S(\text{int}_i) }  \!\!\!\!\! \bq_{\tau,s} -  \!\!\!\!\!\sum_{\tau\in T(\text{int}_i) }  \!\!\!\!\!\bq_{\tau,t}\right),\label{sgenloop}
\end{align}
where we identified $n=0$ and $S(\text{int}_i)$, $T(\text{int}_i)$ denote, respectively, the set of edges starting and ending at the vertex $s(\lambda_{i+1})=t(\lambda_{i})$ and between $\lambda_i$ and $\lambda_{i+1}$ with respect to the ordering. The factor $\varepsilon_i$ in \eqref{sgenloop} is $\varepsilon_i=1$ if $O(\lambda_{i+1},s)>O(\lambda_{i},t)$ (i.~e.~the cilium at $t(\lambda_i)=s(\lambda_{i+1})$ points to the left with respect to the direction of the loop) and $\varepsilon_i=-1$ if $O(\lambda_{i+1},s)<O(\lambda_{i},t)$ (i.~e.~the cilium at $t(\lambda_i)=s(\lambda_{i+1})$ points to the right with respect to the direction of the loop).
 The two edges $\lambda_i$, $\lambda_{i-1}$ are included in these sets if and only if their relative ordering at $s(\lambda_{i+1})=t(\lambda_i)$ changes with respect to the previous vertex.

These quantities are visualised in Fig.~\ref{jloop}. Edges, which transform trivially and which therefore do not contribute to \eqref{angmomgenloop}, \eqref{sgenloop} correspond to brown dashed lines. The edges $\lambda_1,\ldots,\lambda_n$ in the loop are solid black lines. Edges in $S(\text{int}_i)\cup T(\text{int}_i)$ are depicted in red if $\varepsilon_i=-1$ and in blue for $\varepsilon_i=1$. The green edges in Fig.~\ref{jloop} are the ones which contribute to $\bel_\ell$. 
 
Setting $u_\ell=e^{p^a_\ell J_a}$, we find again that the operator $\bel_\ell$ is orthogonal to $\bp_\ell$
while the total spin or internal angular momentum of the loop takes the form
$s_\ell= \tfrac{1}{m}\bp_\ell\cdot \bs_\ell$. 
Following  the discussion in  Sect.~\ref{jqrel}, we can view the vectors $\bq_{\tau,s}$, $\bq_{\tau, t}$ as position vectors of the edge $\tau$ shifted towards its starting and target vertex. This implies that the terms 
 \begin{align}
\bs_i=\Ad(u_{\lambda_{i}}^\inv\cdots u_{\lambda_1}^\inv)(  \sum_{\tau\in S(\text{int}_i) } \!\!\!\!\! \bq_{\tau,s} -\!\!\!\!\!\sum_{\tau\in T(\text{int}_i) }  \!\!\!\!\!\bq_{\tau,t})
 \end{align}
in \eqref{sgenloop} have the interpretation of a relative position vector of the two edges $\lambda_{i+1}$, $\lambda_{i}$  expressed in the reference frame associated with the starting vertex $v=s(\lambda_1)=t(\lambda_n)$ of the loop. The projection of this relative position into the direction of $\bp_{\ell}$ therefore describes an internal angle associated with the vertex $s(\lambda_{i+1})=t(\lambda_{i})$. The total angle associated with the loop which generalises the deficit angles arising in particle spacetimes 
is obtained by summing over the internal  angles of  all vertices in the loop.  In this sum, one has to take into account their  relative position (to the left or right) with respect  to the orientation of the loop which is given by the factors $\varepsilon_i$.

We conclude this section with the discussion of a concrete example based on Fig. \ref{jloopex}.
\begin{example}

We consider a loop $\ell$ such as the one  depicted in Fig. \ref{jloopex} whose
 cilia at the different vertices are chosen such that the ordering is given by: 
\begin{align}
O(\lambda_1,s)<O(\lambda_6,t), ...,\,O(\lambda_2, s)<O(\lambda_1,t),...,\,O(\lambda_k,s)<O(\lambda_{k-1},t),...,\,O(\lambda_6,s)<O(\lambda_5,t)\nonumber.
\end{align}

\begin{figure}[h]
\centering
\psfrag{l1}{$\lambda_1$}
\psfrag{l2}{$\lambda_2$}
\psfrag{l3}{$\lambda_3$}
\psfrag{l4}{$\lambda_4$}
\psfrag{l5}{$\lambda_5$}
\psfrag{l6}{$\lambda_6$}
\psfrag{a}{$\alpha$}
\psfrag{t}{$\tau$}
\psfrag{a}{$\alpha$}
\psfrag{e}{$\eta$}
\psfrag{k}{$\kappa$}
\psfrag{g}{$\gamma$}
\psfrag{b}{$\beta$}
\psfrag{v}{$v$}
      \includegraphics[scale=0.6]{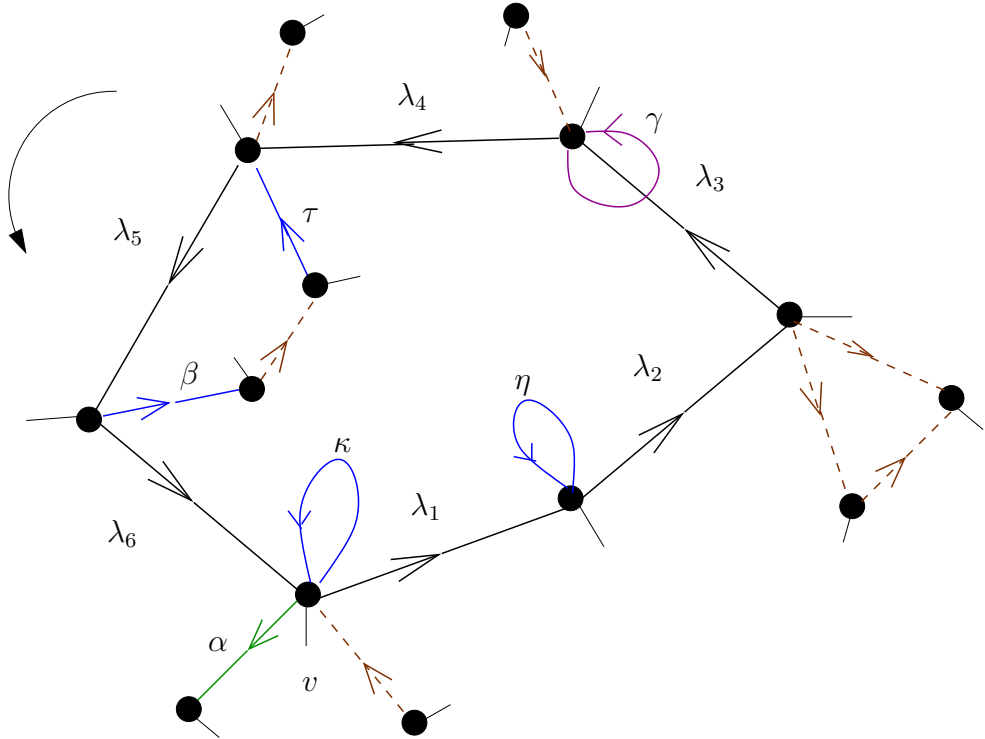}
  \caption{\small{Example of a loop $\ell=\lambda_6\lambda_5\lambda_4\lambda_3\lambda_2\lambda_1$ (black edges) . The variables associated with brown dashed edges transform trivially under $\rho_\ell$. The transformation of the other edge variables is given in \eqref{ledge} and \eqref{otheredge} to \eqref{endother}.}}
  \label{jloopex}
\end{figure}

Expressions \eqref{rhol1act}, \eqref{rhol2act} then imply that the group action $\rho_\ell$ acts on the holonomies of  edges in the loop according to
\begin{align}
\label{ledge}
\rho_\ell(g):\;  
&u_{\lambda_6}\mapsto [g,u_{\lambda_6}u_{\lambda_5}u_{\lambda_4}u_{\lambda_3}u_{\lambda_2}u_{\lambda_1}] \cdot u_{\lambda_6}\\
&u_{\lambda_k}\mapsto u_{\lambda_k}\quad k=1,\ldots,5.\nonumber
\end{align}
For the other edges depicted in Fig.~\ref{jloopex},  the transformation of the holonomies under the group action $\rho_\ell(g)$ is given by \eqref{genedgeact}, \eqref{genedgeact3}, and we obtain
\begin{align}
\label{otheredge}
\rho_\ell(g):\;&u_\alpha\mapsto u_\alpha\cdot [u_{\lambda_6}u_{\lambda_5}u_{\lambda_4}u_{\lambda_3}u_{\lambda_2}u_{\lambda_1},g]\\
&u_\kappa\mapsto g\cdot u_\kappa\cdot g^\inv\\
& u_\eta\mapsto (u_{\lambda_1} g u_{\lambda_1}^\inv)\cdot u_\eta \cdot (u_{\lambda_1}g^\inv u_{\lambda_1}^\inv)\\
&u_\gamma\mapsto u_\gamma\cdot (u_{\lambda_2} u_{\lambda_1})g^\inv(u_{\lambda_2} u_{\lambda_1})^\inv\\
&u_\tau\mapsto (u_{\lambda_4}u_{\lambda_3}u_{\lambda_2}u_{\lambda_1})g(u_{\lambda_4}u_{\lambda_3}u_{\lambda_2}u_{\lambda_1})^\inv \cdot u_\tau\\
&u_\beta\mapsto u_\beta\cdot (u_{\lambda_5}u_{\lambda_4}u_{\lambda_3}u_{\lambda_2}u_{\lambda_1}) g^\inv(u_{\lambda_5}u_{\lambda_4}u_{\lambda_3}u_{\lambda_2}u_{\lambda_1})^\inv.\label{endother}
\end{align}
\end{example}

\subsection{The quantum double  and the kinematical Hilbert space}

\label{kinhilbquant}

The action of the quantum double $D(G)$ associated with each closed, non-selfintersecting loop in $\Gamma$ on the space of cylindrical functions $\cif(G^{|E_\Gamma|})$ does not induce an action of $D(G)$ on the kinematical Hilbert space.  This is due to the fact that multiplication of kinematical states with general functions of the $G$-holonomy along the loop and composition with the associated group action according to  \eqref{doublerep2} does not map kinematical states to kinematical states. However, the kinematical Hilbert space inherits a remnant of these quantum group symmetries which corresponds  to a subalgebra of $D(G)$ generated by two sets of elements.

The first are elements of the form  $f\otimes \delta_e \in D(G)$,  where $f$ is conjugation invariant. They act on the kinematical states by multiplication
\begin{align}
\label{conjinvrep}
\Pi_\ell(f\otimes\delta_e)\psi=f(u_\ell) \cdot \psi.
\end{align}
Since any conjugation invariant function of the $G$-valued holonomy $u_\ell$ is a function of its trace, they are functions of the mass operator  $m_\ell^2=\bp_\ell^2$ which acts according to
\begin{align}
\label{massrep}
\Pi(m_\ell^2)\psi=\bp_\ell^2\cdot \psi.
\end{align}
The second are powers of the ribbon element  \eqref{randc}, which act on cylindrical functions as  \begin{align}
\label{ribbact}
\Pi_\ell(c^k) \psi=\psi\circ\rho_\ell(u_\ell^{-k})\qquad \forall k\in\ZZ.
\end{align}
To demonstrate that the action of these elements does map kinematical states to kinematical states, we note that the group action $\rho_\ell$ satisfies 
\begin{align}
\label{ribbid}
\rho_\ell(h_vu_\ell^t h_v^\inv) \circ G_h=G_h\circ \rho_\ell(u_\ell^t)\quad \forall t\in\RR, h=(h_1,\ldots, h_{|V_\Gamma|})\in G^{|V_\Gamma|}
\end{align}
where $G_h: G^{|E_\Gamma|} \rightarrow G^{|E_\Gamma|}$ is the graph gauge transformation \eqref{ggtrafo} defined by $h$, $h_v$ is the component  of $h$ associated to the starting and target vertex $v$ of $u_\ell$ and
 $u_\ell^t=e^{tp^a_\ell J_a}$.
This identity can be verified by direct calculation for each of the cases considered in Sect.~\ref{explquact}: One considers the action of graph gauge transformations on the edges that share a vertex with the loop and sets  $g=u_\ell^t$ in \eqref{rhol1act} to \eqref{genedgeact3}. Applying this identity to \eqref{ribbact}, one then deduces that the action $\Pi_\ell(c^k)$ commutes with the graph gauge transformations
and hence maps kinematical states to kinematical states.
\begin{align}
(\Pi_\ell(c^k)\psi)\circ G_h=\Pi_\ell(c^k)(\psi\circ G_h)= \Pi_\ell(c^k)\psi\quad\forall\psi\in H_{kin}.
\end{align} 
Moreover, one finds that this action of the ribbon is intimately related to the operator $\bs_\ell$ defined in \eqref{sgenloop} which encodes the internal angular momentum of the loop. Using the results from Sect.~\ref{explquact}, in particular the discussion after \eqref{sgenloop}, we find that the total internal angular momentum of the loop acts on the kinematical Hilbert space via the  infinitesimal version of the action \eqref{ribbact}
\begin{align}
\label{spinrep}
\Pi(m_\ell s_\ell)\psi=\Pi(\bp_\ell\bj_\ell)\psi=i\frac{d}{dt}|_{t=0}\psi\circ \rho_\ell(u_\ell^t).
\end{align}
Hence, for each closed, non-selfintersecting loop in the graph $\Gamma$, the associated action of the quantum double $D(G)$ on the space of cylindrical functions gives rise to two sets of operators acting on the kinematical Hilbert space: the mass operator $m_\ell$ which acts by multiplication and the product $m_\ell s_\ell$ of mass and spin which acts via the group action $\rho_\ell$. As discussed in the previous sections, these are the two fundamental physical observables associated to each loop $\ell$ in the graph. They correspond to the two Casimir operators of the three-dimensional Euclidean and Poincar\'e group and have a clear physical interpretation through the analogy with the corresponding variables for particles.

\section{Construction of the physical Hilbert space}

\label{physhilb}
In this section, we discuss the implementation of the remaining constraints and the construction of the physical Hilbert space in the loop and the combinatorial formalism. As exhibited in the previous sections, the absence of local gravitational degrees of freedom implies
that no refinements of the graphs are required to capture the local geometry of the spacetime. 
The Hamiltonian constraint therefore a priori does not act by adding edges around vertices as in the four-dimensional case. Instead, it takes the form of a flatness condition $F_\ell\approx0$
 \eqref{discrete constraints} on the graph connections, which requires the $G$-valued holonomy around each contractible loop in the graph to be trivial.

\subsection{The physical Hilbert space in the loop formalism}

\label{loophilb}

The construction of the physical Hilbert space of three-dimensional loop quantum gravity 
has been investigated extensively as a toy model for the four-dimensional case \cite{AHRSS, SL, AA, MD, ARL}. For reasons of simplicity, much of the previous work in this context focussed on 
 the Euclidean case with a torus as the spatial surface $S$. Here, we adopt the
presentation given in \cite{AK} which is more general and presents a convenient starting point for the comparison with the combinatorial
quantisation formalism.

In \cite{AK}, the discrete version of the flatness constraint $F_\ell$  \eqref{discrete constraints} is implemented by means of a   
``projector"  $P:H_{kin}\rightarrow H_{phys}$ on the physical Hilbert space\footnote{Although this map does not have the property $P\circ P=P$ associated with the notion of a projector, we refer to it as ``projector" in the following, since this is the prevalent convention in the literature.}. Formally, this projector  acts on the kinematical states associated with a graph $\Gamma$ according to \begin{align}
\label{physproj}
P:\; \psi\mapsto \!\!\!\!\!\ \!\!\!\!\! \prod_{   \substack{{\ell \;\text{closed,}}   \\{\text{contractible loop in}\;\Gamma}  }}\!\!\!\!\!\ \!\!\!\!\!\delta_e(u_\ell)\;\;\cdot\; \psi\qquad\forall \psi \in H_{kin}^\Gamma
\end{align}
where $u_\ell=u_{\lambda_n}\cdots u_{\lambda_1}$ is the $G$-holonomy along the contractible loop $\ell=\lambda_n\circ\ldots\circ \lambda_1$ and the product runs over all contractible loops in $\Gamma$. As this expression involves a product of delta-distributions,
it is a priori ill-defined and requires a regularisation.  
In the case $G=SU(2)$, a regularisation scheme was proposed and lead to an explicit relation between the Ponzano-Regge
model and three dimensional loop quantum gravity \cite{AK}. 
Given a suitable regularisation of the projector $P$, one can construct the physical Hilbert space $H_{phys}$ as the image of the kinematical Hilbert space  ${H}_{kin}$ under $P$ up to zero norm vectors.
Identifying these zero norm vectors amounts to identifying gauge equivalent states or gauge fixing the symmetries which are 
generated by the curvature constraint $F(x)=0$.

In practice, this gauge fixing procedure proceeds in two steps. The first is to remove most of the redundant degrees of freedom encoded  in the kinematical states $H_{kin}^\Gamma$ by contracting the underlying graph $\Gamma$ along a maximal connected tree \cite{AK} (see \cite{LPR} for a detailed discussion in the spin-foam approach).  This results in a graph,
with only one vertex and with edges that are loops based at that vertex, which we will refer to as ``flower graph" in the following.
The second step is to remove the residual gauge degrees of freedom associated with  the flower graph by imposing the flatness condition on each contractible loop and by imposing the mass and spin constraint for each loop around a particle. For the details of this procedure we refer the reader to \cite{NP,K1,AK}, for a  discussion in the context of spin foam models see also \cite{LPR}. In the following we will focus on the general picture and its relation to the combinatorial approach.

\subsection{The physical Hilbert space in the combinatorial formalism}
\label{hphyscomb}

In the combinatorial formalism, 
 the implementation of the constraint $F=0$ is intimately related to the representations of the quantum double $D(G)$  in Sect.~\ref{qudoublesymms} and their remnants on the kinematical Hilbert space  in Sect.~\ref{kinhilbquant}.  To understand this point, we recall the formula \eqref{conjinvrep} for the representation of the quantum double associated with the loop 
$\ell$. Applying this formula to the delta-distribution  $\delta_{C_\mu}$ on the space of $G$-conjugacy classes $C_\mu$, we find 
 \begin{align}
\Pi_\ell (\delta_{C_{\mu}}\otimes \delta_e)\psi= \delta_{C_\mu}(u_\ell)\cdot \psi.\end{align}
In the case $G=SU(2)$ the  conjugacy classes are labelled by an angle $\mu \in [0,2\pi]$ and the
unitary irreducible representations by a spin $J=\tfrac 1 2, 1,\ldots$. The delta-distribution can then be realised as the familiar sum over the characters $\chi_J$ as follows
\begin{align}
\delta_{C_\mu}(u_\ell)=\sum_{J=\tfrac 1 2, 1 ,\tfrac 3 2,\ldots} \chi_J(u_\ell) \chi_J(e^{\mu J_0}).\end{align}
In the case $G=SU(1,1)$  the situation is more complicated due to its non-compactness. However, we note that in both cases the restriction to the fixed conjugacy class implemented by this delta-distribution projects on the space of eigenstates of the mass operator $m_\ell^2$ \eqref{massrep} with eigenvalue $\mu^2$, i.~e.~on the subspace  of kinematical states satisfying
\begin{align}
\Pi(m_\ell^2)\psi=\mu^2\cdot \psi.
\end{align}
The projector \eqref{physproj} on the physical Hilbert space corresponds to projecting on states
for which the holonomy along $\ell$ is trivial. It is therefore 
 implemented by the remnant of the quantum double representations on the kinematical Hilbert space 
\begin{align}
P:\;\psi\mapsto \!\!\!\!\!\ \!\!\!\!\!\prod_{   \substack{{\ell \;\text{closed,}}   \\{\text{contractible loop in}\;\Gamma}  }}\!\!\!\!\!\!\!\!\Pi_\ell (\delta_{C_0}\otimes \delta_e)\;\;\cdot\;\; \psi,
\end{align}
where $C_0=\{e\}$ is the conjugacy class containing the identity element. 
The other kinematical operator associated with a closed loop in the graph is the ribbon element which acts via \eqref{ribbact} and  corresponds to the product of mass and spin \eqref{spinrep}. Imposing invariance under the action of these kinematical operators amounts to requiring that the kinematical states are invariant under the associated one-parameter group of transformations $\rho_\ell(u_\ell^{-t})$ for all contractible loops $\ell$ in $\Gamma$ or, equivalently, that the product of its mass and spin vanishes
\begin{align}
\Pi_\ell(c^t)\psi=\psi\circ\rho_\ell(u_\ell^{-t})= \psi\quad\forall t\in\RR\quad \Leftrightarrow\quad \Pi(m_\ell s_\ell)\psi=0.
\end{align}

The constraints associated with loops around particles are implemented analogously, only that in this case the group elements are restricted to a fixed conjugacy class with $\mu\neq 0$, such that the projector implementing this condition is given by 
\begin{align}
\label{partmass}
\psi\mapsto \Pi_\ell(\delta_{C_\mu}\otimes\delta_e)\,\cdot \psi.\end{align}
Similarly, the states are no longer required to be invariant under the associated group action $\rho_\ell$ but to transform  covariantly, i.~e.~to be eigenstates of the 
 operator $m_\ell s_\ell$ with eigenvalue $\mu s$, where $s$ is the spin of the particle
 \begin{align}
 \label{spinpart}
 \pi_\ell(c^t)\psi=e^{it\mu s}\cdot \psi\qquad\Leftrightarrow\qquad \Pi(m_\ell s_\ell)\psi=\mu s\cdot\psi.
\end{align}

\subsection{Gauge fixing via contracting a maximal tree and graph contractions}

After discussing the general formalism for the imposition of the constraints in the loop and in the combinatorial formalism, we will now focus on the two steps in its practical implementation, the gauge fixing procedure via contractions of maximal trees and the imposition of the residual constraints on the resulting flower algebra. In  this subsection, we demonstrate
 that the gauge fixing procedure via contraction of a maximal tree in the graph $\Gamma$ is intimately related to the graph operations in Fock and Rosly's description of the phase space \cite{FR} and their quantum counterparts.
 
\begin{figure}[h]
\centering
\psfrag{l}{$\lambda$}
\psfrag{S}{$S(\lambda)$}
\psfrag{T}{$T(\lambda)$}
\psfrag{G}{$\Gamma$}
\psfrag{Gp}{$\Gamma'$}
      \includegraphics[scale=0.6]{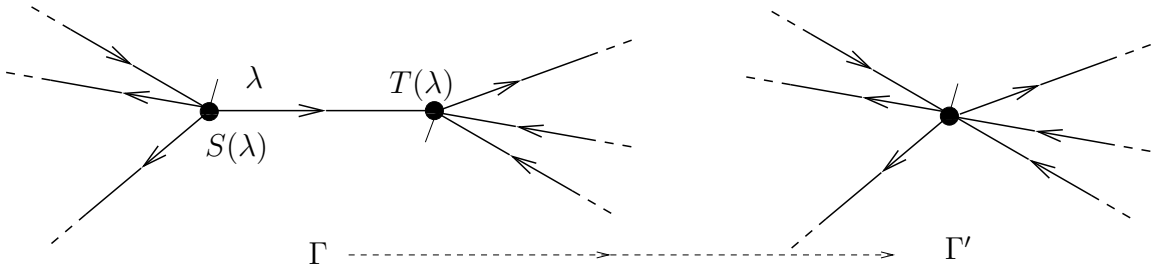}
  \caption{\small{Illustration of the contraction operation of a graph $\Gamma$ to a graph $\Gamma'$. 
The contraction reduces the number of vertices and edges of the graph by one.}}
  \label{contraction}
\end{figure}

We start by outlining the notion of graph contractions as defined in \cite{FR}. 
Given a graph $\Gamma$ and an edge $\lambda\in E_\Gamma$, one can contract $\lambda$ either towards its starting point or endpoint. Contracting the edge $\lambda$ towards the starting vertex $s(\lambda)$ amounts to performing a gauge transformation at its endpoint $t(\lambda)$ that sets the group element $H_\lambda=(u_\lambda,-\Ad(u_\lambda)\bj_\lambda)$ equals to one, removing the edge  $\lambda$ and the cilium at $t(\lambda)$ and inserting all edges incident at $t(\lambda)$ at the former starting point of $\lambda$ as shown in Fig.\ref{contraction}.  Contraction towards the target vertex is defined analogously. 
The result is a graph $\Gamma'$ with $|E_{\Gamma'}|=|E_\Gamma|-1$ edges and $|V_{\Gamma'}|=|V_\Gamma|-1$ vertices. 

From \eqref{ggtrafo} it follows that this procedure introduces a map $\Phi_\lambda: IG^{|E_\Gamma|}\rightarrow IG^{|E_\Gamma|-1}$ between the  $IG$-valued holonomies associated to the edges of the graphs $\Gamma$, $\Gamma'$. For the contraction towards the starting vertex of $\lambda$, it acts on the
$IG$-valued holonomies $H_\tau$, $\tau\in\Gamma\setminus{\lambda}$ according to
\begin{align}\label{scontract}
\Phi_\lambda:\; 
&H_\tau\mapsto H_\tau'=\begin{cases} H_\lambda^\inv\cdot H_\tau  &\text{for}\; \tau\in T(t(\lambda))\\
  H_\tau\cdot H_\lambda & \;\text{for}\;
 \tau\in S(t(\lambda))\\
 H_\tau & \text{otherwise}\end{cases}
\end{align}
 The corresponding map for contraction towards the starting vertex is obtained by replacing $t(\lambda)$ by $s(\lambda)$ in \eqref{scontract} and by exchanging left-multiplication by $H_\lambda$ and right-multiplication with $H_\lambda^\inv$. 

The map \eqref{scontract}  commutes with the graph gauge transformations \eqref{ggtrafo}  in the following sense: Consider the graph gauge transformation $G_h^\Gamma$ for $\Gamma$ defined by an element $h=(h_1,\ldots, h_{|V_\Gamma|})\in IG^{|V_\Gamma|}$. Denote by $h' $  the element  of  $IG^{|V_\Gamma|-1}$ obtained by omitting the  entry $h_{t(\lambda)}$ for the target vertex of $\lambda$ and by $G^{\Gamma'}_{h'}$ the associated graph gauge transformation for $\Gamma'$. Then
\begin{align}
\label{gtrafocommute}
G^{\Gamma'}_{h'}\circ \Phi_\lambda=\Phi_\lambda\circ G^{\Gamma}_h.
\end{align}
Moreover, it has been shown by Fock and Rosly \cite{FR} that the map \eqref{scontract} is a Poisson map between the spaces of graph connections associated to $\Gamma$ and $\Gamma'$, i.e.~that it maps the Poisson structure for the graph $\Gamma$ to the one for $\Gamma'$ 
\begin{align}
\label{Poissmap}
\{ f\circ\Phi_\lambda, g\circ \Phi_\lambda\}_{\Gamma}=\{f,g\}_{\Gamma'}\circ\Phi_\lambda\qquad \forall f,g\in\cif(IG)^{|E_{\Gamma}|-1}
\end{align}
Due to the close link between classical and quantum theory apparent in \eqref{jbhkin},  these results translate immediately into analogous statements for operators acting on the cylindrical functions and the associated kinematical Hilbert spaces. 

We start by considering the cylindrical functions associated to the graphs $\Gamma$ and $\Gamma'$. Via its restriction  $\phi_\lambda: G^{|E_{\Gamma}|} \rightarrow G^{|E_{\Gamma}|-1}$ to the $G$-components of the holonomies, $\Phi_\lambda$ induces a map from the space of cylindrical functions  for $\Gamma'$ to the space of cylindrical functions for $\Gamma$ which acts on cylindrical states as follows
\begin{align}
\psi_{\Gamma'}\mapsto \psi_{\Gamma'}\circ\phi_\lambda.
 \end{align}
The fact that $\Phi_\lambda$ is a Poisson map then implies via \eqref{jact} that the representations of the operators $\bj_\tau$, $\tau\in \Gamma$ and the operators $\bj_{\tau'}$, $\tau'\in\Gamma'$ are compatible in the following sense
\begin{align}
\left(\Pi_{\Gamma'}(j^a_{\tau'})\psi_{\Gamma'}\right)\circ \phi_\lambda=\Pi_\Gamma(j^a(H'_\tau))(\psi_{\Gamma'}\circ \phi_\lambda)\qquad\forall \psi_{\Gamma'}\in \cif(G^{|E_\Gamma|-1})
\end{align}
where $\bj(H'_\tau)$ is the angular momentum vector of the holonomy $H'_\tau$ given by \eqref{scontract}
\begin{align}
\bj(H'_\tau)=\begin{cases} \bj_\tau-\Ad(u_\tau^\inv u_\lambda)\bj_\lambda & \tau \in T(t(\lambda))\\
\Ad(u_\lambda^\inv)\bj_\tau+\bj_\lambda & \tau\in S(t(\lambda))\\
\bj_\tau & \text{otherwise}.
\end{cases}
\end{align}
Similarly, we have for the representation of functions associated with $\Gamma,\Gamma'$
\begin{align}
\left(\Pi_{\Gamma'}(f_{\Gamma'})\psi_{\Gamma'}\right)\circ\phi_\lambda=\Pi_{\Gamma}(f_{\Gamma'}\circ \phi_\lambda)(\psi_{\Gamma'}\circ \phi_\lambda)\qquad \forall \psi_{\Gamma'}\in\cif(G^{|E_\Gamma|-1})
\end{align}
Hence, contracting an edge towards a vertex induces an homomorphism from the algebra of quantum operators acting on the cylindrical functions for  $\Gamma'$ to the algebra of quantum operators acting on the cylindrical functions for $\Gamma$.

We will now demonstrate that these graph contractions give rise to an isomorphism of the kinematical Hilbert spaces $H_{kin}^\Gamma$, $H_{kin}^{\Gamma'}$ with the corresponding inner products. 
For this, we note that \eqref{gtrafocommute} implies that the map $\Phi_\lambda$ preserves invariance under graph gauge transformations and hence induces a map
\begin{align}
\phi_\lambda^{kin}:\; H_{kin}^{\Gamma'}\rightarrow H_{kin}^\Gamma\qquad \psi_{\Gamma'}\mapsto \psi_{\Gamma'}\circ \phi_\lambda\in H^\Gamma_{kin}.
\end{align}
To show that this map is an isomorphism, we need to define its inverse.  For this we introduce a map 
$\Xi_\lambda:\;IG^{|E_{\Gamma}|-1}\rightarrow IG^{|E_\Gamma|}$
 which inserts the identity element for the holonomy of the the contracted edge $\lambda$
\begin{align}
\label{xidef}
\Xi_\lambda:\;& 
(H_1,\ldots, H_{|E_\Gamma|-1})\mapsto (H_1,\ldots, 1,\ldots, H_{|E_\Gamma|-1} ),
\end{align}
and denote by $\xi_\lambda: G^{|E_{\Gamma}|-1}\rightarrow G^{|E_\Gamma|}$ the associated map acting on the $G$-valued holonomies. To show that $\Xi_\lambda$ commutes with graph gauge transformations and satisfies a relation analogous to \eqref{gtrafocommute}, we consider 
 a  general graph gauge transformation for $\Gamma'$ given by an element
$h=(h_1,\ldots, h_{|V_\Gamma|}-1)\in IG^{|V_{\Gamma}|-1}$. We denote by $h_\lambda$ the entry associated to the vertex obtained by contracting $\lambda$.  
We define the associated element  $h' \in IG^{|V_\Gamma|}$ by inserting the entry $h_\lambda$ for both the arguments $s(\lambda)$ and $t(\lambda)$. It then  follows from the definition \eqref{xidef} of $\Xi_\lambda$ and \eqref {ggtrafo} that the associated  graph gauge transformations $G^{\Gamma'}_h$ $G^\Gamma_{h'}$ satisfy a relation analogous to \eqref{gtrafocommute} and thus preserve graph gauge invariance
\begin{align}
\label{gtrafocommute2}
\Xi_\lambda\circ G^{\Gamma'}_h = G^\Gamma_{h'}\circ\Xi_\lambda.
\end{align}
They therefore induce a  map between the associated Hilbert spaces  $H_{kin}^\Gamma$, $H_{kin}^{\Gamma'}$ 
\begin{align}
\xi_\lambda^{kin}:\; H_{kin}^{\Gamma}\rightarrow H_{kin}^{\Gamma'}\qquad \psi_{\Gamma}\mapsto \psi_{\Gamma}\circ \xi_\lambda\in H^{\Gamma'}_{kin},
\end{align}
and  \eqref{scontract}, \eqref{xidef} imply $\Phi_\lambda\circ\Xi_\lambda=1$. This proves that the maps $\phi_\lambda^{kin}$, $\xi_\lambda^{kin}$ are isomorphisms from $H^{kin}_{\Gamma'}$ to $H^{kin}_\Gamma$ and vice versa. 
The condition \eqref{Poissmap}, which states that graph contractions are Poisson maps, ensures that the  action of the kinematical observables associated with the edges of $\Gamma$, $\Gamma'$  on  $H^{kin}_{\Gamma'}$, $H^{kin}_\Gamma$ are obtained as the images of the corresponding actions on  $H^{kin}_{\Gamma}$, $H^{kin}_{\Gamma'}$.

Moreover,  it follows directly from the definition of the maps \eqref{scontract}, \eqref{xidef} that these isomorphisms preserve the scalar product \eqref{physscalprod}. Applying a graph gauge transformation analogous to the one in \eqref{scontract} that sets the group element $u_\lambda$ to one and using the 
 left-and right invariance of the Haar measure on $G$, one obtains after a redefinition of the integration variables
\begin{align}
&\langle  \psi_{\Gamma'}\circ\phi_\lambda, \chi_{\Gamma'}\circ\phi_\lambda \rangle_{\Gamma}=\int_{G^{|E_\Gamma|}} d\mu(u_1,..., u_{|E_\Gamma|})\;\; \overline{\psi_{\Gamma'}}\circ\phi_\lambda (u_1,..., u_{|E_\Gamma|})\chi_{\Gamma'}\circ\phi_\lambda(u_1,..., u_{|E_\Gamma|})\nonumber\\
&=\text{vol}(G)\int_{|G^{|E_\Gamma|-1}} \!d\mu(u_1,..., \widehat{u_\lambda},..., u_{|E_\Gamma|}) \;\;\overline{ \psi}_{\Gamma'} (u_1,..., \widehat{u_\lambda},..., u_{|E_\Gamma|})\chi_{\Gamma'}(u_1,..., \widehat{u_\lambda},..., u_{|E_\Gamma|})\nonumber\\
&=\text{vol}(G)\cdot \langle  \psi_{\Gamma'}, \chi_{\Gamma'} \rangle_{\Gamma'},
\end{align}
where $\widehat{u_\lambda}$ denotes omission of the argument associated to the edge $\lambda$. For the case $G=SU(2)$, this reflects the familiar invariance of the Ashtekar-Lewandowski measure in the context of loop quantum gravity while the expression diverges for $G=SU(1,1)$. This is due to the non-compactness of $SU(1,1)$ and demonstrates an additional need for gauge fixing in the non-compact case. 

\begin{figure}[h]
\centering
      \includegraphics[scale=0.6]{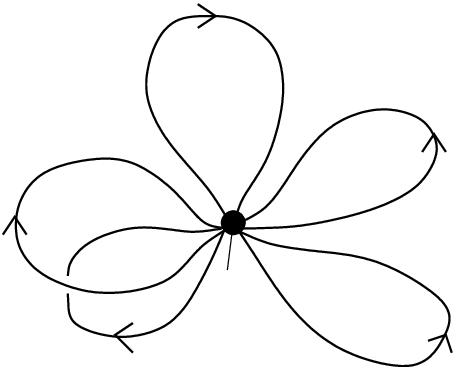}
  \caption{\small{A flower graph consisting of a single vertex and edges that are loops attached to the vertex. Depending on the topology of the underlying surface, each loop can be either contractible
or non-contractible. }}
  \label{flower}
\end{figure}

By selecting a maximal tree in the graph $\Gamma$ and repeatedly applying the contraction procedure to the edges of this tree, one obtains a flower graph with a single vertex and edges that are loops as  depicted in Fig. \ref{flower}.
Hence, the familiar gauge fixing procedure in the loop formalism via contraction of trees has a  canonical interpretation in the combinatorial formalism based on the description of the phase space of Fock and Rosly \cite{FR}. It arises as the quantum counterpart of the edge contractions on the phase 
space of the theory which act on the  $IG$-holonomies of the edges. The $G$-component of these graph contractions defines their action on the  cylindrical functions and the kinematical states associated with the graphs $\Gamma$, $\Gamma'$.
Their translational part relates the operators $\bj_\tau$, $\bj_{\tau'}$ for edges $\tau\in \Gamma$, $\tau'\in\Gamma'$ and the corresponding kinematical operators. The fact that edge contractions are Poisson maps \cite{FR} ensures that the action of the operators $\bj_{\tau'}$, $\tau'\in \Gamma'$ is obtained as the image of the action of $\bj_\tau$, $\tau\in\Gamma$ and that the action of the kinematical operators for the two graphs commutes with the graph contractions.

\subsection{Residual gauge freedom and the construction of the physical Hilbert space}

\label{flowerconst}

\begin{figure}[h]
\centering
\psfrag{m}{$M_1$}
\psfrag{a1}{$A_1$}
\psfrag{a2}{$A_2$}
\psfrag{b1}{$B_1$}
\psfrag{b2}{$B_2$}
\psfrag{k}{$k$}
      \includegraphics[scale=0.6]{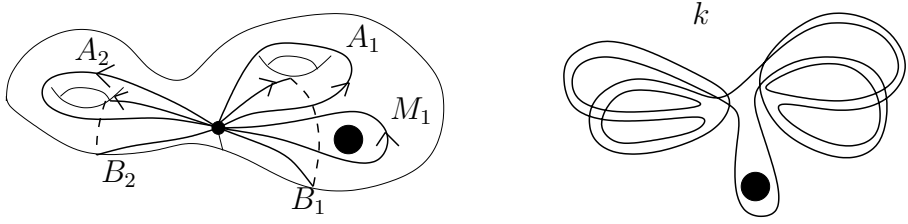}
  \caption{\small{The flower graph associated to a surface $S$ of genus 2 punctured with one particle. 
The loops, denoted $A_1$, $A_2$, $B_1$, $B_2$ and $M$, are in correspondence with the generators of the fundamental group $\pi_1(S)$.  The loop $k$ depicted on the right is defined algebraically in \eqref{defofk}.}}
  \label{fundamental}
\end{figure}

After the contraction of a maximal tree, in both formalism the resulting graph is a flower graph as depicted in Fig.\ref{flower}.
The residual graph gauge transformations act by simultaneous conjugation of the $G$-holonomies associated to all edges, and there are three classes of residual constraints:
\begin{enumerate}
\item A flatness constraint $u_\ell\approx 1$ for each contractible petal; \vspace{-.2cm}
\item A particle constraint which restricts the petals around particles to a fixed conjugacy classes determined by mass and spin of the particle;\vspace{-.2cm}
\item An additional constraint $u_k\approx 1$ implementing the condition that the curve $k$
depicted in the Fig.~(\ref{fundamental}) is contractible.\vspace{-.2cm}
\end{enumerate}
The first set of constraints is implemented by simply removing the contractible petals from the flower graph. The edges of the resulting graph  then define a set of generators of the spatial surface's fundamental group $\pi_1(S)$  as illustrated in 
Fig.~\ref{fundamental}.  For a surface $S$ of genus $g$ with $n$ punctures, this set of generators consists of loops $M_i$, $i=1,\ldots,n$, around each puncture and two curves $A_j,B_j$, $j=1,\ldots,g$ for each handle as shown in Fig.~\ref{fundamental}. It is subject to a single defining relation which amounts to imposing that the curve $k$ in Fig.~\ref{fundamental} is contractible 
 \begin{align}
\label{defofk}
 k=B_g\circ A_g^\inv\circ B_g^\inv\circ A_g\circ... B_1\circ A_1^\inv\circ B_1^\inv\circ A_1\circ M_n\circ...\circ M_1=1.
 \end{align}
The associated cylindrical functions depend on the $G$-holonomies $u_{M_1}$, ... ,  $u_{M_n}$, $u_{A_1}$ , $u_{B_1}$, ... , $u_{A_g}$, $u_{B_g}\in G$ along these generators. The  kinematical states are functions of these $G$-holonomies which are invariant under simultaneous conjugation with $G$
\begin{align}
\label{pi1kin}
H_{\pi_1}^{kin}=\{ \psi\in\cif(G^{n+2g})\;|\; \psi(hu_{M_1}h^\inv,...,hu_{B_g}h^\inv)=\psi(u_{M_1},...,u_{B_g})\}.
\end{align}
The
 Hamiltonian
 constraint then reduces to the requirement that the $G$-holonomy along the curve $k$ in Fig.~\ref{fundamental} vanishes and implements the defining relation of the fundamental group $\pi_1(S)$.
This implies that the projector
 on the physical Hilbert space $H_{phys}$ takes the form
\begin{align}
\label{pi1proj}
P:\;\psi\in H_{kin} \mapsto \delta_e([u_{B_g}, u_{A_g}^\inv]\cdots [u_{B_1}, u_{A_1}^\inv] u_{M_n}\cdots u_{M_1})\cdot \psi.
\end{align}

In the combinatorial formalism, the requirement of graph gauge invariance \eqref{pi1kin} and the  constraint implemented by the projector \eqref{pi1proj} are combined into the requirement that the physical states transform trivially under the representation of the quantum double $D(G)$   associated with the curve $\ell=k$. As shown in \cite{we2}, see in particular Sect.~4.2. there, but also directly apparent from the explicit expressions for the group action in Sect.~\ref{explquact}, this representation acts on the cylindrical functions according to
\begin{align}
\label{constdouble}
\Pi_k(f\otimes \delta_h)\psi(u_{M_1},...,u_{B_g})=f([u_{B_g}, u_{A_g}^\inv]\cdots u_{M_1})\cdot \Psi(hu_{M_1}h^\inv,..., h u_{B_g} h^\inv)
\end{align}
such that the combined action of the Hamiltonian constraint operator \eqref{pi1proj} and the graph gauge transformations takes the form
\begin{align}
\Pi_k(\delta_e\otimes \delta_h)\psi(u_{M_1},..., u_{B_g})=\delta_e([u_{B_g}, u_{A_g}^\inv]\cdots u_{M_1})\cdot \Psi(hu_{M_1}h^\inv,..., h u_{B_g} h^\inv).
\end{align}

The remaining gauge freedom is the one associated to the particle constraints, which are given by the action of the mass and spin operators of the loops around each particle
\begin{align}
\label{pi1partcond}
\Pi(m_i^2)\psi =\mu_i^2 \cdot \psi\qquad \Pi(m_is_i)\psi=\mu_is_i\cdot \psi\qquad i=1,\ldots,n.
\end{align}
The canonical way of implementing these conditions \eqref{pi1partcond} in the combinatorial formulation is 
discussed in \cite{we2}.
It consists in parametrising the corresponding $G$-holonomies as
\begin{align}
&u_{M_i}=v_{M_i}e^{\mu_i J_0} v_{M_i}^\inv \qquad v_{M_i}\in G,\;i=1,\ldots,n
\end{align}
and working with cylindrical functions that depend on the variables  $v_{M_i}$ instead of $u_{M_i}$. 
The implementation of the  spin constraints in \eqref{pi1partcond} is then directly related to the representation theory of the quantum double summarised in appendix \ref{qurep}. Denoting by
$N_{\mu_i}$ the centraliser of the conjugacy class $C_{\mu_i}$ as defined in \eqref{centraliser} and by 
$\pi_{s_i}$  its irreducible unitary representation introduced labelled by $s_i$, one finds that the spin constraints \eqref{pi1partcond} take the form
\begin{align}
&\psi(v_{M_1},...,v_{M_i} n_i,...,u_{A_1},..., u_{B_g}) =\pi_{s_i}(n_i^\inv)\psi(v_{M_1},..., v_{M_n}, u_{A_1},...u_{B_g}) \qquad \forall n_i\in N_{\mu_i}.
\end{align}

Moreover, it is shown in \cite{we2}, that the representation $\Pi_k$ of the quantum double which implements the residual constraints then takes
the form
\begin{align}
\label{finconst}
\Pi_k(f \otimes \delta_h)\psi(v_{M_1},...,v_{M_n}, ..., u_{A_g},u_{B_g})=f(u_k)\cdot \Psi( h v_{M_1},...,hv_{M_n},..., h u_{B_g} h^\inv).
\end{align}
This expression for the action of the Hamiltonian constraint and the graph gauge transformations
establishes a direct link between the construction of the physical Hilbert space of the theory and the representation theory of the quantum double $D(G)$.
Using the formulas  \eqref{irrep of DG}, \eqref{irrep2} for the irreducible representations of the quantum double $D(G)$ and the formula \eqref{adrep} for the adjoint action of $D(G)$ on itself, one can rewrite \eqref{finconst} as
\begin{align}
\label{repid}
\Pi_k(f\otimes \delta_g)\psi\!\! =\!\!\left(\Pi_{\mu_1 s_1}\!\otimes\!...\!\otimes\! \Pi_{\mu_n s_n}\!\otimes\!\text{ad}\!\otimes\! ...\! \otimes\! \text{ad}\right)
\left((\Delta\!\otimes\! 1\!\otimes\!...\!\otimes\! 1)\circ ...\circ (\Delta\!\otimes\! 1)\right)\;\psi,
\end{align}
where $\Delta$ is the coproduct \eqref{coprod} of $D(G)$  \cite{we2}. Hence, the implementation of the constraints is intimately related to the construction of the tensor product of certain irreducible and adjoint representations of the quantum double $D(G)$. 
This is a further manifestation of the role of the quantum double $D(G)$ as a quantum symmetry of the theory and its role in the construction of the physical Hilbert space. Note also that it does not only involve the algebra structure of the quantum double which encodes the underlying Poincar\'e or Euclidean  symmetry of the classical theory but also its coproduct, which differs from the trivial coproduct of the universal enveloping algebras of the three-dimensional Lorentz and Poincar\'e algebras. In this sense, the quantum double $D(G)$ appears naturally as a {\em deformation} of the $IG$-symmetry in the classical theory. 

The presence of quantum double symmetries in the quantum theory is not only of conceptual importance but also provides concrete advantages in the construction of the physical Hilbert space and the quantisation of the theory. Equation \eqref{repid} reduces the implementation of the Hamiltonian constraint and the construction of the physical scalar product to a mathematical problem from the representation theory of the quantum double $D(G)$: It states that the implementation of the constraints amounts to the construction of the invariant subspace in the tensor product of certain representations of $D(G)$. 

For the case $G=SU(2)$, the decomposition of a tensor  product of two irreducible representations of $D(G)$ is given in \cite{KBM}. While the general case and the decomposition for the non-compact group $G=SU(1,1)$ present considerable technical challenges, the link between the implementation of the constraints and the quantum double $D(G)$ makes the construction of the physical Hilbert space amenable to techniques from the representation theory of quantum groups. In particular, it provides a canonical set of physical states in the framework of representation theory, namely  the characters of the quantum double. For the case of Chern-Simons theory with gauge group $SL(2,\CC)$ which corresponds to Lorentzian and Euclidean 3d gravity with, respectively, positive and negative cosmological constant,  these states have been constructed and investigated in \cite{BNR}. For the case of vanishing cosmological constant, these physical states are constructed in
 \cite{MN2}.

\section{Outlook and Conclusions}

\label{concsect}

In this paper we clarified the relation between three-dimensional loop quantum gravity and the combinatorial quantisation formalism based on the Chern-Simons formulation of the theory.
We related the construction of the kinematical and physical Hilbert space in the two approaches and established an explicit relation between the associated quantum operators.
Although the (extended) Hilbert spaces in the two formulations are identical, the basic operators acting on these spaces differ in the two approaches. While the operators in the loop formalism are defined generically, the definition of the operators in the combinatorial formalism requires an additional structure associated with the graph. This additional structure is a ciliation, which defines a linear ordering of the incident edges at each vertex and enters already in the description of the classical theory  \cite{FR}. 

This ciliation manifests itself also in the explicit relation between these operators, which we derived in this paper, and in their physical interpretation: The operators in the loop formalism can be viewed as position vectors for the edges with respect to a fixed reference frame. In contrast, the operators in the combinatorial formalism correspond to a relative position vector of two edge ends with respect to a reference frame associated with its starting vertex. Defining this relative position vector requires the choice of a reference point at each edge or, equivalently, the choice of a ciliation.  
In the case of edges which are loops, the corresponding combinatorial operator gives rise to an internal angle variable  
and an external reference angle associated with the loop. In this case, the ciliation is required to establish the notion of ``internal"  and ``external'' and to define the corresponding angles. 

The second core result of our paper is our clarification of the role of quantum group symmetries, more specifically the quantum doubles $D(SU(2))$, $D(SU(1,1))$, in the two formalisms. We showed that these symmetries are present naturally also in the loop formalism: Each closed non-selfintersecting loop in the graph gives rise to a representation of the quantum double on the space of cylindrical functions. The explicit expressions for these representations,  which we derived in this paper, depend again on the choice of a ciliation. 
This result demonstrates that quantum group symmetries are a generic feature of three-dimensional quantum gravity with vanishing cosmological constant which are also present in the loop formalism.  Moreover, we showed that they play an important role in the implementation of the constraints and the construction of the physical Hilbert space.
The explicit determination of the physical states will be investigated in \cite{MN2} for the case where the spatial surface
is a torus.

While our results clarify the relation between three-dimensional loop quantum gravity and the combinatorial quantisation formalism as well as the role of quantum group symmetries in the theory, many other aspects remain to be investigated.
Specifically, it would be interesting to determine how our results are related to the constraint implementation in 
\cite{TT3D} and to the work \cite{Ltalk, LZart}.
The former studies the implementation of constraint by adding edges around each vertex as in the four-dimensional case.
The latter is also concerned with the relation between quantisation approaches based on the Chern-Simons formulation  and  quantisation approaches based on the BF formulation of three-dimensional gravity. However, it appears that the basic variables investigated in this work are different and quantum group symmetries are not apparent there.

It would be also instructive to investigate the relation between the combinatorial quantisation formalism and other quantisation approaches for three-dimensional gravity with a non-vanishing cosmological constant.
However, we expect these cases to be more subtle. The direct relation between the Hilbert space in the combinatorial formalism and cylindrical and spin network functions  based on the groups $SU(2)$, $SU(1,1)$ for vanishing cosmological constant is a consequence of the semidirect product structure
of the associated symmetry groups. Generically, quantum states are constructed from the irreducible representations of the associated quantum groups. For non-vanishing cosmological constant, the relevant quantum groups are not the quantum doubles of groups but  the quantum doubles of $q$-deformed universal enveloping algebras whose Hopf algebra structure and representation theory are more involved. 

In the loop formalism, the cosmological constant does a priori not affect the construction of the kinematical Hilbert space and  enters the formalism only in the implementation of the Hamiltonian constraint.  Hence, if quantum group symmetries are present in the loop formalism for non-vanishing cosmological constant, their emergence should be the result of the implementation of the Hamiltonian
constraint.  It would be very interesting to understand if and how such quantum group symmetries arise. 
A preliminary study of this question is given in 
 \cite{Ale}, but many issues remain to be clarified.  
It can therefore be anticipated that  the relation between combinatorial quantisation, loop quantum gravity and spinfoam models  will be less direct for non-vanishing cosmological constant.

\section*{Acknowledgements}
The research of K.N. was partially
supported by the ANR (BLAN06-3\_139436 LQG-2006). The research of C.M. was supported by  the DFG Emmy-Noether research grant ME 3425/1-1 and by the Perimeter Institute for Theoretical
Physics.  Research at Perimeter Institute is supported by the Government
of Canada through Industry Canada and by the Province of Ontario through
the Ministry of Research \& Innovation. C.M. thanks the Laboratoire de Math\'ematiques et de Physique Th\'eorique in Tours
 for their hospitality. This work was initiated and partially completed during her visits in Tours which were also supported
by the ANR.

\appendix

\section{Fock and Rosly's Poisson structure in Chern-Simons theory and  three-dimensional gravity}
\label{frphsp}

In this appendix, we summarise  Fock and Rosly's description \cite{FR} of the phase space 
of Chern-Simons theory and its application to three-dimensional gravity with vanishing cosmological constant. We start by considering the formalism for a general Chern-Simons theory with gauge group $H$ and denote by $\gothh$ the associated Lie algebra.

The two central ingredients in Fock and Rosly's description of the phase space are
 an oriented graph $\Gamma$ with a cilium added at each vertex as explained in Sect.~\ref{graphsect} and a classical $r$-matrix for the group $H$ which is compatible with the Chern-Simons action. The latter is an element $r\in\gothh\otimes\gothh$  which satisfies the following two conditions: 
\begin{enumerate}
\item It is  a solution of the classical Yang Baxter equation 
\begin{align}
\label{cybe}
&[[r,r]] \equiv [r_{12}, r_{13}]+[r_{12}, r_{23}]+[r_{13}, r_{23}]=0\\
&r_{12}\equiv r^{\alpha \beta} \xi_\alpha\otimes \xi_\beta \otimes 1, 
\quad r_{13}:=r^{\alpha \beta} \xi_\alpha\otimes 1\otimes \xi_\beta,
\quad r_{23}:=r^{\alpha\beta} 1\otimes \xi_\alpha\otimes \xi_\beta,\nonumber
\end{align}
where $r=r^{\alpha \beta}\, \xi_\alpha\otimes \xi_\beta$ is the expression for $r$ in a fixed basis
$\{\xi_\alpha\}_{\alpha=1,\ldots,\text{dim}\,\gothh}$ of the Lie algebra $\gothh=\text{Lie}\,H$.

\item  Its symmetric part  $r_s=\tfrac{1}{2} (r^{\alpha \beta}+r^{\beta\alpha}) 
\xi_\alpha\otimes \xi_\beta$
is dual to the $Ad$-invariant symmetric form $\langle\,,\,\rangle$ in the Chern-Simons action or, in other words, it is given by the associated Casimir operator of $\gothh$.
\end{enumerate}
It has been shown by Fock and Rosly that, together with a ciliated graph $\Gamma$ as in Sect.~\ref{graphsect}, such classical $r$-matrices define a Poisson structure on the manifold $H^{|E_\Gamma|}$.
The different copies of $H$ correspond to the $H$-valued holonomies obtained by integrating the gauge field along the edges of $\Gamma$, and after imposition of  the discretised  flatness constraints, the Poisson structure agrees with the canonical symplectic structure on the moduli space of flat $H$-connections modulo gauge transformations.

Fock and Rosly's Poisson structure is most easily expressed in terms of a Poisson bivector
\begin{align}
\label{pb}
\{F,G\}=(dF\otimes dG)(B_{FR})\qquad\forall F,G\in\cif(H),
\end{align}
which takes the form
\begin{align}
\label{bivec}
B_{FR}=\sum_{v\in V_\Gamma} r^{\alpha\beta}(v)&\bigg(\tfrac 1 2\sum_{\lambda\in S(v)}  \xi_\alpha^{R,\lambda}\wedge \xi_\beta^{R,\lambda}  +\tfrac 1 2 \sum_{\lambda\in T(v)} \xi_\alpha^{L,\lambda}\wedge \xi_\beta^{L,\lambda}\\
+&\!\!\sum_{\lambda\in S(v)} \xi_\alpha^{R,\lambda}\wedge \big(\!\!\!\!\!\sum_{\tau\in S^+(s(\lambda))}  \!\!\!\!\!\xi_\beta^{R,\tau}+\!\!\!\!\!\!\sum_{\tau\in T^+(s(\lambda))} \!\!\!\!\!\xi_\beta^{L,\tau}\big)+\!\!\!\sum_{\lambda\in T(v)} \xi_\alpha^{L,\lambda}\wedge \big(\!\!\!\!\!\sum_{\tau\in S^+(t(\lambda))}  \!\!\!\!\!\xi_\beta^{R,\tau}+\!\!\!\!\!\sum_{\tau\in T^+(t(\lambda))} \!\!\!\!\!\xi_\beta^{L,\tau}\big).\nonumber
\bigg)
\end{align}
Here, $r^{\alpha\beta}(v)$ stands for components of the classical $r$-matrices assigned to the vertices of the graph and satisfying the two conditions above\footnote{As these conditions do not necessarily define the $r$-matrix uniquely, different $r$-matrices can be assigned to different vertices as long as they satisfy these conditions.}.  All notations referring to the graph $\Gamma$ are defined as in Sect.~\ref{graphsect}, and $\xi_\alpha^{L,\lambda}$, $\xi^{R,\lambda}_\beta$ denote the right- and left-invariant vector fields associated to the basis elements $\xi_\alpha\in\gothh$ and the different copies of $H$. Their action on 
functions $F\in\cif(H^{|E_\Gamma|})$ is given by
\begin{align}
&\xi_\alpha^{L,\lambda}F(h_1,\ldots, h_{|E_\Gamma|})=\frac d {dt}|_{t=0} F(h_1,\ldots, e^{-t\xi_\alpha} \cdot h_{\lambda},\ldots,h_{|E_\Gamma|})\\
&\xi_\alpha^{R,\lambda}F(h_1,\ldots, h_{|E_\Gamma|})=\frac d {dt}|_{t=0} F(h_1,\ldots, h_\lambda\cdot e^{t\xi_\alpha} ,\ldots,h_{|E_\Gamma|})\nonumber.
\end{align}

We are now ready to discuss the application of Fock and Rosly's description to three-dimensional gravity with vanishing cosmological constant.  In this case, we have $H=IG$, and the associated Lie algebras $\gothh$ are the three-dimensional Euclidean and Poincar\'e algebra with generators $\{\xi_\alpha\}=\{J_a,P_a\}_{a=0,1,2}$ and Lie bracket \eqref{poincbrack}.  It has been shown in \cite{BMS, we1, K1}  that the relevant classical $r$-matrix  for  the Chern-Simons formulation of three-dimensional gravity takes the form
\begin{align}
\label{rmat}
r=P_a\otimes J^a.
\end{align}
To derive an expression for Fock and Rosly's Poisson structure in terms of functions $f\in\cif (G^{|E_\Gamma|})$ of the $G$-valued holonomies $u_\lambda$ and the vectors $\bj_\lambda$ associated to the edges $\lambda\in E_\Gamma$, one needs to determine the action of the right- and left-invariant vector fields $J_a^{R,\lambda}$, $J_a^{L,\lambda}$, $P_a^{R,\lambda}$, $P_a^{L,\lambda}$ 
on these variables. This has been done in \cite{BMS,we1,we2}, but can also be inferred directly from their definition and the group multiplication law \eqref{gmult}. With the notations introduced above, one  finds that their action on functions $f\in\cif (G^{|E_\Gamma|})$ is given by
\begin{align}
\label{vfu1}
&J^a_{L, \lambda} f(u_1,\ldots, u_{|E_\Gamma|})=L^a_{\lambda} f(u_1,\ldots, u_{|E_\Gamma|})=\frac d {dt}|_{t=0}  f(u_1,\ldots, e^{-tJ_a}\cdot u_\lambda,\ldots, u_{|E_\Gamma|})\\
&R^a_{L, \lambda} f(u_1,\ldots, u_{|E_\Gamma|})=R^a_{\lambda} f(u_1,\ldots, u_{|E_\Gamma|})=\frac d {dt}|_{t=0}  f(u_1,\ldots, e^{-tJ_a}\cdot u_\lambda,\ldots, u_{|E_\Gamma|})\\
&P^a_{L,\lambda}f(u_1,\ldots, u_{|E_\Gamma|})=P^a_{R,\lambda}f(u_1,\ldots, u_{|E_\Gamma|})=0,
\end{align}
and that their action on the variables
variables $j_\tau^a$, $\tau\in E_\Gamma$, takes the form
\begin{align}
\label{poincvfj}
&J^a_{L, \lambda}\, j_\tau^b=0 &  &J^a_{R,\lambda} \,j^b_\tau=\delta_{\lambda,\tau}\,\epsilon^{ab}_{\;\;\;\;c}\, j^c_\lambda\\
&P^a_{L,\lambda}\,j_\tau^b=-\delta_{\lambda,\tau}\, \eta^{ab} &  &P^a_{R,\lambda}\,j^b_\tau=\delta_{\lambda,\tau}\,\Ad(u_\lambda)^{ab},\label{vfj}
\end{align}
where $\delta_{\lambda,\tau}=1$ if $\tau=\lambda$ and vanishes otherwise.
By inserting the classical $r$-matrix \eqref{rmat} and expressions \eqref{vfu1} to \eqref{vfj} into the general formulae \eqref{pb}, \eqref{bivec}, one then obtains the Poisson brackets of the variables $\bj_\lambda$, $\lambda\in E_\Gamma$ and $f\in\cif(G^{|E_\Gamma|})$. As the vector fields $P^a_{L,\lambda}$, $P^a_{R,\lambda}$ act trivially on functions $f\in\cif(G^{|E_\Gamma|})$, the Poisson bracket of the latter vanishes
\begin{align}
\{f,g\}=0\qquad \forall f,g \in\cif(G^{|E_\Gamma|}).
\end{align}
A short calculation shows that the Poisson brackets of the variables $j^a_\lambda$ with functions $f\in\cif(G^{|E_\Gamma|})$ are given by \eqref{jbrack}
\begin{align}
\label{jbrackapp}
\{j^a_\lambda, f\}=  -R_\lambda^a f-\!\!\!\!\!\!\!\!\!\sum_{\tau\in S^+(s(\lambda))}\! \!\!\!\!\!\!R_\tau^af -\!\!\!\!\!\!\sum_{\tau\in T^+(s(\lambda))} \!\!\!\!\!\!\!L_\tau^af 
+\Ad(u_\lambda^\inv)^{a}_{\;\;b}\left(\sum_{\tau\in S^+(t(\lambda))} \!\!\!\!\!\!\!R_\tau^bf +\!\!\!\!\!\!\sum_{\tau\in T^+(t(\lambda))} \!\!\!\!\!\!\!L_\tau^bf \right) .
\end{align}
This implies that one can identify the variables $j^a_\lambda$ with certain vector fields $X^a_\lambda$ on $G^{|E_\Gamma|}$ and that their Poisson brackets are given by the Lie brackets of these vector fields
via
\begin{align}
\{\{j^a_\lambda,j^b_\tau\}, f\}=[X_\lambda^a,X^b_\beta] f\qquad\forall f\in\cif(G^{|E_\Gamma|}).
\end{align}

\section{The representation theory of the quantum double $D(G)$}

\label{qurep}
In this appendix, we give a brief summary of the representation theory of the quantum double $D(G)$. For a detailed treatment we refer the reader to \cite{kM, KBM}.

We start by recalling the observation that the quantum double $D(G)$ is a Drinfeld deformation of the group algebra
$\mathbb C(IG)$  and that 
$D(G)$ is included into $\mathbb C(IG)$ as an algebra. This  inclusion  of $\CC(IG)$ into $D(G)$ implies that the  irreducible unitary representations of $D(G)$ give rise to representations to the three-dimensional Poincar\'e and Euclidean group $IG$. The latter are 
labelled  by two parameters  $(\mu,s)$ where the $\mu$ is a real number usually interpreted as a mass and
 $s$ is an integer when $G=SU(2)$ or a real number when $G=SU(1,1)$ and stands for an internal angular momentum or spin. the products
$\mu^2$ and $\mu s$ are, respectively, the eigenvalues of the Casimir $P^2$ and $P\cdot J+J\cdot P$ in the associated representations of the Lie algebras \eqref{poincbrack}.
Hence, the mass $\mu$ defines a  $G$-conjugacy class $\mathcal C_\mu$ 
 and the spin $s$ an irreducible representation $\pi_s: N_\mu\rightarrow \text{End}(V_s)$  of its centraliser
\begin{align}
\label{centraliser}
N_\mu=\{n\in G\;|\; ngn^\inv=g\quad\forall g\in\mathcal C_\mu \}.
\end{align}
In the case where $G=SU(2)$,  conjugacy classes $\mu$ are angles in the interval $[0,2\pi[$. 
The centralisers $N_\mu$ are isomorphic to the group $U(1)$ when $\mu >0$
and to $G$ otherwise. Generically (when $\mu>0$), representations $\pi_s$ of the centraliser are 
therefore labelled by an integer $s$.

 The Hilbert spaces of the the representations $(\mu,s)$ are
\begin{align}
\label{doublerep}
V_{\mu s}= \{ &\psi:G\rightarrow V_{s} \;|\; \psi(vn) =
\pi_s (n^{-1})
\psi(v),\;\;\forall n\in N_\mu,\; \;\forall v \in G,\nonumber\\
&\mbox{and} \;
\|\psi\|^2:= \int_{G/N_\mu} \|\psi(z)\|_{V_s}^2\,dm(zN_\mu)<\infty\}/\sim,
\end{align}
where $\sim$ denotes division by zero-norm states and $dm$ is an
invariant
measure on $G/N_\mu$. The quantum double $D(G)$ acts on these  spaces according to
\begin{align}\label{irrep of DG}
\Pi_{\mu s} (F)\psi(v)=\int_G d\mu(z) F(vg_\mu v^\inv, z)\psi(z^\inv v),
\end{align}
where $g_\mu$ is a fixed element of the conjugacy class $\mathcal C_\mu$ and $d\mu(z)$ denotes the Haar measure on $G$.  For the  singular elements $f\otimes\delta_g$ this expression simplifies to
\begin{align}
\label{irrep2}
\Pi_{\mu s}( f\otimes \delta_g)\psi (v)= f(vg_\mu v^{-1}) \psi(g^{-1}v).
\end{align}
Another representation which plays an important role in the quantisation of three-dimensional gravity is the adjoint representation obtained by letting
$D(G)$ act on itself via the adjoint action
\begin{align}
\label{adrep}
&\ad(F) \phi (w_1,w_2)= \int_Gd\mu(z)\,\,F(w_1w_2^{-1}w_1^{-1}w_2,z) \phi(z^{-1}w_1z,z^{-1}w_2z)\qquad F,\phi\in D(G)\\
\label{singdoublerep}
&\ad( f\otimes\delta_g)
\phi (w_1,w_2)
= f(w_1w_2^{-1}w_1^{-1}w_2) \phi(g^{-1}w_1g,g^{-1}w_2 g)\,.
\end{align}
As an illustration, let us consider once again the example $G=SU(2)$. In that case, the vector space $V_{\mu s}$
is simply $\{f \in F(G) \vert f(xh(\theta)) = e^{is\theta} f(x),\, \forall \theta \in [0,2\pi[, \, x\in G\}$ 
where $h(\theta)$ is the diagonal representative of the conjugacy class $\theta$. The Hilbert structure is
given by  the Haar measure of $SU(2)$. The action of $D(SU(2))$ can be deduced immediately from 
(\ref{irrep of DG}).  Of particular relevance are the representations of the ribbon element \eqref{randc} and of the character $\chi$ in the fundamental representation of $SU(2)$, which are diagonal and can be viewed as the "deformed" (or exponentiated) version of the classical Casimir elements of 
the Euclidean algebra 
\be
\Pi_{\mu s}(c) \psi (v) \; = \; e^{-i\mu s} \cdot \psi(v) \;\;\;\;\; \text{and} \;\;\;\;\;
\Pi_{\mu s} (\chi \otimes 1)\psi(v) \; = \; 2 \cos \mu \cdot \psi(v) \;. 
\ee

\end{document}